\documentclass[a4paper,12pt]{article}
\usepackage[a4paper,text={16.8cm,22.4cm}]{geometry}
\usepackage{amsmath,amssymb,bm,graphicx,color,psfrag,booktabs,textcomp,multirow}
\usepackage[normalem]{ulem}
\usepackage{cancel}
\allowdisplaybreaks 
\def\rd{\mathrm{d}}

\def\rG{\mathrm{\Gamma}}
\def\bI{\boldsymbol{I}}
\def\bT{\boldsymbol{T}}
\def\sip{\hat{\sigma}}
\def\up{\hat{u}}
\def\tp{\hat{t}}

\def\qb{\bar{q}}

\def\eps{\epsilon}

\renewcommand{\Re}[1]{{\rm Re} #1}
\newcommand*{\dsl}[1]{\cancel{#1}} 

\begin{document}

\begin{titlepage}

\begin{flushright}
September 12, 2013\\
\hfill OUTP-13-19P
\end{flushright}

\vspace{0.7cm}
\begin{center}
\Large\bf
\boldmath
Transverse-momentum spectra of electroweak bosons \\ near threshold at NNLO
\unboldmath
\end{center}

\vspace{0.2cm}
\begin{center}
{\sc Thomas Becher$^a$, Guido Bell$^b$, Christian Lorentzen$^a$ and Stefanie Marti$^a$}\\
\vspace{0.4cm}
{\sl 
 ${}^a$\,Albert Einstein Center for Fundamental Physics\\
Institute for Theoretical Physics \\
University of Bern\\
Sidlerstrasse 5, 3012 Bern, Switzerland\\[0.4cm]

${}^b$\,Rudolf Peierls Centre for Theoretical Physics\\ 
University of Oxford\\ 
1 Keble Road, Oxford OX1 3NP, United Kingdom}
\end{center}

\vspace{1.0cm}
\begin{abstract}
\vspace{0.2cm}
\noindent 
We obtain the next-to-next-to-leading order corrections to transverse-momentum spectra of $W$, $Z$ and Higgs bosons near the partonic threshold. In the threshold limit, the electroweak boson recoils against a low-mass jet and all radiation is either soft, or collinear to the jet or the beam directions. We extract the virtual corrections from known results for the relevant two-loop four-point amplitudes and combine them with the soft and collinear two-loop functions as defined in Soft-Collinear Effective Theory. We have implemented these results in a public code {\sc PeTeR} and present numerical results for the threshold resummed cross section of $W$ and $Z$ bosons at next-to-next-to-next-to-leading logarithmic accuracy, matched to next-to-leading fixed-order perturbation theory. The two-loop corrections lead to a moderate increase in the cross section and reduce the scale uncertainty by about a factor of two. The corrections are significantly larger for Higgs production.
\end{abstract}

\vfil

\end{titlepage}

\section{Introduction}

It is often believed that the virtual corrections are the most complicated piece of higher-order perturbative computations, but in general this is not true. Beyond next-to-leading order (NLO), the real-emission corrections typically present the main difficulty. For example, because of the complicated structure of the real emissions, the NNLO corrections to $W$, $Z$ and photon production in association with an energetic jet are currently still unknown, despite the fact that the two-loop virtual corrections to these processes have been available for more than ten years \cite{Garland:2001tf,Garland:2002ak}.  Over the past years, a number of techniques have been developed to isolate the soft and collinear singularities at NNLO so that the remaining phase-space integrations can be performed numerically, see e.g.\ \cite{Anastasiou:2003gr,GehrmannDeRidder:2005cm,Czakon:2010td}. Using these methods, the NNLO corrections for dijet production \cite{Ridder:2013mf} and for Higgs-boson production in association with a jet \cite{Boughezal:2013uia} have recently been obtained for the dominant, purely gluonic, partonic channel. While these computations are very challenging, we can  expect that NNLO results for all electroweak bosons will become available in the not too distant future.

Soft and collinear singularities complicate the numerical evaluation of phase-space integrals, but the amplitudes themselves greatly simplify in the singular regions. In the present paper we will use this simplification to obtain an approximate NNLO result for Higgs-boson, $W$, $Z$ and photon production in association with an energetic jet. To do so, we will consider the production of an electroweak boson $V$ near the partonic threshold. Near threshold, the electroweak boson recoils against a low-mass hadronic jet, and the real radiation is either soft, or collinear to the jet or the incoming partons. As a consequence, the partonic cross section in the channel $a+b \to V+ j_c $ factorizes
\begin{equation} \label{eq:factorization}
\hat{s} \frac{\rd\hat{\sigma} }{\rd\hat{u}\, \rd\hat{t}} = \hat{\sigma}^{(0)}_{ab}( \hat{u},\hat{t},\mu)\, 
\hat{H}_{ab} (\hat{u},\hat{t},\mu)
 \int\! \rd k\, J_c (m_X^2-2 E_J k,\mu) \,S_{ab}(k,\mu)\, ,
\end{equation} 
where $\hat{\sigma}^{(0)}_{ab}$ is the Born cross section and
the partonic Mandelstam variables are $\hat{s} = (p_a+p_b)^2$, $\hat{t} = (p_a-q)^2$ and $\hat{u} = (p_b-q)^2$, with $q$ the vector-boson momentum,
and $q^2=M_V^2$. The jet functions $J_q$ and $J_g$ describe the collinear radiation initiated by an energetic quark or gluon, respectively. The two-loop quark jet function  was computed in \cite{Becher:2006qw}, and the two-loop gluon result was obtained in \cite{Becher:2010pd}. Last year, also the two-loop soft function $S_{ab}$ was computed \cite{Becher:2012za}. Here, we will determine the final NNLO ingredient, the two-loop hard function $\hat H_{ab}$, for all partonic channels for both vector-boson and Higgs-boson production. The hard function contains the virtual corrections which were computed in \cite{Garland:2001tf,Garland:2002ak,Gehrmann:2011ab,Gehrmann:2013vga,Gehrmann:2011aa}. We will convert these results into an infrared-subtraction scheme that is compatible with the jet and soft function calculations. Our paper completes the construction of the two-loop cross section near the partonic threshold. Since the threshold terms usually amount to the bulk of the cross section, we expect that our result is a good approximation to the full NNLO result. 

The derivation of the factorization formula (\ref{eq:factorization}) in Soft-Collinear Effective Theory (SCET) \cite{Bauer:2000yr,Bauer:2001yt,Beneke:2002ph} was given in \cite{Becher:2009th} for photon production and generalized to the $W$ and $Z$ case in \cite{Becher:2011fc,Becher:2012xr}. An interesting new feature, which first arises at NNLO is the presence of a purely gluonic channel $gg\to V  g $ in $Z$-boson and photon production. This was already pointed out in \cite{Becher:2009th}, but here we explicitly give the relevant SCET operators and determine their Wilson coefficients using the results \cite{Gehrmann:2013vga} for the corresponding loop amplitudes. Numerically, however, we find that this channel only gives a negligible contribution. The factorization formula (\ref{eq:factorization}) can also be used to resum the threshold terms to all orders. For $W$ and $Z$ production at large transverse momentum, this resummation was performed at next-to-leading logarithmic (NLL) accuracy in~\cite{Kidonakis:1999ur,Kidonakis:2003xm,Gonsalves:2005ng} and to NNLL in~\cite{Becher:2009th,Becher:2011fc,Kidonakis:2012sy,Kidonakis:2011hm}. With all two-loop ingredients in place, we can extent the resummation to N$^3$LL accuracy since the necessary anomalous dimensions are known. We have implemented the N$^3$LL  resummation, as well as the NNLO fixed-order expansion of the threshold cross section into a public code {\sc PeTeR} ("large-$p_T$ Resummation'') \cite{peter}. Numerically, we find that the resummation effects are small and the N$^3$LL resummed results are close to the NNLO threshold results. 

Our paper is organized as follows. In Section \ref{sec:hard} we show how to extract the Wilson coefficients of the SCET operators describing $V+{\rm jet}$ production from the results for four-point amplitudes in the literature. The construction of the associated hard functions is then detailed in Section \ref{sec:helicity}. We first treat the vector-boson case, where we discuss the $q\bar{q}\rightarrow V g$ and $qg\rightarrow Vq$ channels in Section  \ref{sec:quark} and the $gg \to V g$ process in Section \ref{sec:gluon}. After this, we construct the hard functions for Higgs production in Section \ref{sec:higgs}. With all the ingredients in place, we then compute the two-loop threshold cross section in Section \ref{sec:expansion} and study the numerical impact of the corrections in Section \ref{sec:numerics}. 

\section{On-shell matching and renormalization\label{sec:hard}}

The hard functions in SCET are obtained by performing a matching calculation, i.e.\ by computing the same quantity in QCD and in the effective theory, and then fixing the Wilson coefficients of the SCET operators in such a way that the QCD result is reproduced. The Wilson coefficients are independent of the process and the external states used to perform the matching. By far the simplest possibility is to use on-shell amplitudes, in our case $q\bar{q}\rightarrow V g$, $qg\rightarrow Vq$ and $gg\rightarrow Vg$, since the loop integrals in the effective theory are scaleless for on-shell momenta and vanish in dimensional regularization. The relevant two-loop QCD amplitudes were obtained in \cite{Garland:2002ak,Gehrmann:2011ab,Gehrmann:2013vga,Gehrmann:2011aa}. We now explain how the SCET operators are constructed and how their bare Wilson coefficients are obtained from QCD results for the on-shell amplitudes, and then perform the renormalization of these coefficients. 

The SCET operators mediating the production of an electroweak boson at large transverse momentum $p_T$ involve collinear fields associated with the two beam directions and the direction of the associated jet. At leading power in the effective theory, they involve a single collinear field for each of the three directions. To construct the collinear Lagrangians, one introduces a light-cone reference vector for each direction. The vectors $n_1$ and $n_2$ point in the beam directions, while $n_J$ is along the jet direction. Each reference vector $n_i$ has a conjugate light-cone vector $\bar{n}_i$, with $n_i\cdot \bar{n}_i=2$. Quarks collinear to the direction $i$ are described by a field $\chi_i$, which fulfills the condition $n_i\!\!\!\!\!/\,\,\chi_i=0$, so that this field is effectively a two-component field. Also, at leading power, only the components of the gluon field ${\cal A}_{i\perp}^{\nu,a}$ transverse to its light-cone direction can contribute. Because of these conditions,  there is a one-to-one correspondence between helicity states and associated operators. Helicity amplitudes are therefore particularly well suited to extract the SCET matching coefficients, as stressed in \cite{Becher:2009cu,Becher:2009qa,Stewart:2012yh}.

We will now explain the relation between helicity amplitudes and Wilson coefficients in detail, using the example of the  purely gluonic channel which arises for $Z$ and $\gamma$ production at one-loop and contributes at NNLO to the cross section. The presence of SCET operators mediating the $gg\to V g$ process was pointed out in \cite{Becher:2009th}, but in contrast to the operators for $q\bar{q}\rightarrow V g$ and $qg\rightarrow Vq$, the purely gluonic operators were not explicitly given since they are not needed at NNLL accuracy. According to our considerations from above, the leading-power operators for $gg\to V g$ (and for $gg\to H g$)  have the form
\begin{align}\label{eq:ggg}
	\mathcal{O}_{abc}^{\nu\rho\sigma} (x; t_1,t_2,t_J) &= 
	{\cal A}_{1\perp}^{\nu,a}(x+ t_1 \bar{n}_1) \,{\cal A}_{2\perp}^{\rho,b}(x+ t_2 \bar{n}_2) \,
	{\cal A}_{J\perp}^{\sigma,c}(x+ t_J \bar{n}_J) \, .
\end{align}
The operators for the quark channels such as  $q\bar{q}\rightarrow V g$ have the same structure, but involve a quark field, an anti-quark field and a gluon. In SCET the collinear operators are smeared over the directions associated with the large external momenta, and the associated hadronic vector current $\mathcal{J}_\mu(x)$ mediating $gg\to V g$  consists of a convolution of the Wilson coefficients with the smeared operators
\begin{equation}
	\mathcal{J}_\mu(x) =  \int d t_1\, d t_2 \, d t_J \,
	C^{abc}_{\mu\nu\rho\sigma}(t_1,t_2,t_J)\, \mathcal{O}_{abc}^{\nu\rho\sigma}(x; t_1,t_2,t_J) \,.
\end{equation}
Because we have left the color and Lorentz structure of the fields in the operators open, the Wilson coefficients have Lorentz and color indices, too. These are contracted with the operators to ensure that $\mathcal{J}^\mu(x) $ transforms as a singlet under color and a vector (or axial-vector) under the Lorentz group. The color structure of the Wilson coefficients can be either symmetric or antisymmetric. In the first case, it is proportional to $d^{abc}$, in the second case proportional to the structure constants $f^{abc}$. Working with open Lorentz and color indices is convenient because the coefficients $C^{abc}_{\mu\nu\rho\sigma}$ are directly related to helicity amplitudes in color space. To see this, we perform the matching using on-shell $gg \to V g$ amplitudes. Since the loop integrals in the effective theory are all scaleless for on-shell external momenta,
the effective theory amplitudes reduce to tree-level matrix elements multiplied by the Wilson coefficients. Furthermore, because the different collinear sectors no longer interact after soft-collinear decoupling, the matrix element factorizes into individual collinear matrix elements, which in a given sector have the form
 \begin{equation}
 \langle 0|\, {{\cal A}^{\nu,a}_{j\perp}}(t_j\bar n_j)\, |p_i;a_i,\lambda_i
   \rangle
  = \delta_{ij}\,\delta_{a_i a}\,e^{-it_i\bar n_i\cdot p_i}\,
    \epsilon^\nu(p_i,\lambda_i) \,
\end{equation}
for an incoming transverse gluon field. Performing the integrations over the variables $t_i$, one then finds that the Fourier transforms of the Wilson coefficients $C^{abc}_{\mu\nu\rho\sigma}(t_1,t_2,t_J)$, contracted with the external polarization vectors, are equal to the helicity amplitudes. 

The vanishing of the loop corrections in the effective theory implies that in the relevant integrals, the infrared (IR) and ultraviolet (UV) singularities exactly cancel each other. Since the IR singularities of QCD and SCET are the same, this further implies that the UV singularities of SCET Wilson coefficients are identical to the IR singularities of QCD amplitudes. As a consequence the IR singularities of $n$-point amplitudes  in $d=4-2\epsilon$ dimensions can be renormalized multiplicatively 
\cite{Becher:2009cu,Becher:2009qa}
\begin{align}\label{zren}
|\mathcal{M}^{\text{ren}}(\{p\},\,\mu)\rangle=\lim_{\epsilon\rightarrow 0}\,\boldsymbol{Z}^{-1}(\epsilon,\,\{p\},\,\mu)\,|\mathcal{M}(\epsilon,\,\{p\})\rangle \,,
\end{align}
where the renormalization factor $\boldsymbol{Z}$ is a matrix in color space. It is spin-independent, but depends logarithmically on the external  momenta $\{p \}\equiv{p_1,\dots,\,p_n}$. The renormalized amplitude $|\mathcal{M}^{\text{ren}}(\{p\},\,\mu)\rangle$ is equal to the renormalized Wilson coefficient of the leading-power SCET operator with the same quantum numbers as the external states in the amplitude. The inverse of the $\boldsymbol{Z}$-factor can be written in terms of the anomalous dimension matrix \cite{Becher:2009cu}
\begin{align}
\boldsymbol{\rG}(\{p\},\mu)=\sum_{(i,j)}\frac{\bT_i\cdot \bT_j}{2}\,\gamma_{\text{cusp}}(\alpha_s) \ln\frac{\mu^2}{-s_{ij}}+\sum_i\gamma^i(\alpha_s) \, ,
\label{eq:hardanomdim}
\end{align}
where the sum runs over unordered tuples $(i,j)$ of distinct partons, $\bT_i$ is the color generator associated with the $i$-th parton in the scattering amplitude, which acts on the color index of that parton, and $s_{ij}\equiv2\sigma_{ij} p_i\cdot p_j+i0$, where the sign factor $\sigma_{ij}=+1$ if the momenta $p_i$ and $p_j$ are both incoming or outgoing, and $\sigma_{ij}=-1$ otherwise. The product $\bT_i\cdot\bT_j\equiv T_i^a\,T_j^a$ is summed over $a$. Generators associated with different particles trivially commute, $\bT_i\cdot \bT_j=\bT_j\cdot \bT_i$ for $i\neq j$, and $\bT_i^2=C_i$ is given in terms of the quadratic Casimir operator of the corresponding color representation, i.e.~$C_q=C_{\bar{q}}=C_F$ and $C_g=C_A$. For more details on the color-space formalism, see \cite{Catani:1996jh,Catani:1996vz}.

In \cite{Garland:2002ak,Gehrmann:2011ab,Gehrmann:2013vga,Gehrmann:2011aa} the results were given in terms of finite helicity amplitudes obtained after removing the IR singularities using Catani's subtraction formula \cite{Catani:1998bh}. In the following, we will relate these expressions to the renormalized SCET Wilson coefficients. The entire procedure can be viewed as a scheme change from Catani's subtraction scheme to a standard $\overline{\text{MS}}$ subtraction of the singularities.

\subsection{Conversion to $\overline{\text{MS}}$ scheme}

We first reconstruct the IR-divergent part of the two-loop amplitudes and will then perform the renormalization. We write the expansion of the UV-renormalized, on-shell $n$-parton scattering amplitude with IR singularities regularized in $d=4-2\epsilon$ dimensions as
\begin{align}
|\mathcal{M}(\epsilon,\{p\})\rangle\equiv \mathcal{M}^{(0)}+\frac{\alpha_s}{2\pi} \mathcal{M}^{(1)}(\eps)
+\left(\frac{\alpha_s}{2\pi}\right)^2\mathcal{M}^{(2)}(\eps)
+\mathcal{O}(\alpha_s^3) \,,
\label{Mdiv}
\end{align}
where $\alpha_s\equiv\alpha_s(\mu)$ is the renormalized coupling constant. Note that the superscript $(i)$ refers in this section to an expansion in units of $\alpha_s/2\pi$, which is the notation adopted in the literature on two-loop four-point functions \cite{Garland:2002ak,Gehrmann:2011ab,Gehrmann:2013vga,Gehrmann:2011aa}. In the SCET literature, the perturbative expansion is usually written in $\alpha_s/4\pi$.  Throughout this section, we will expand in $\alpha_s/2\pi$ to be compatible with the literature on the amplitudes, but we will switch to the standard SCET notation when we present our result for the cross section in Section \ref{sec:expansion} and in the appendices.

The helicity amplitudes in \cite{Garland:2002ak,Gehrmann:2011ab,Gehrmann:2013vga,Gehrmann:2011aa} were constructed using Catani's IR-subtraction formula \cite{Catani:1998bh}, which states that the product
\begin{align}
|\mathcal{M}^{\text{fin}}(\{p\},\mu)\rangle &=   \left[1-\frac{\alpha_s}{2\pi}\,\boldsymbol{I}^{(1)}(\epsilon)-\left( \frac{\alpha_s}{2\pi} \right)^2 \boldsymbol{I}^{(2)}(\epsilon) +\dots \right] |\mathcal{M}(\epsilon,\{p\})\rangle \,
\end{align}
is free of IR poles through $\mathcal{O}(\alpha_s^2)$. The amplitudes are however different from the $\overline{\text{MS}}$-renormalized amplitudes $|\mathcal{M}^{\text{ren}}(\{p\},\mu)\rangle$ in (\ref{zren}), because the subtraction operators 
$\boldsymbol{I}^{(n)}(\epsilon)\equiv \boldsymbol{I}^{(n)}(\epsilon,\,\{p\},\,\mu)$ 
contain terms of arbitrarily high orders in $\epsilon$. The explicit form of the $\boldsymbol{I}^{(n)}(\epsilon)$ can be found in Appendix \ref{renapp}. The above relation can be inverted to reconstruct the expansion coefficients of the IR-divergent amplitude $|\mathcal{M}(\epsilon,\{p\})\rangle$ as
\begin{align}
\mathcal{M}^{(1)}(\eps)
&= \mathcal{M}^{(1),\,\text{fin}}+\bI^{(1)}(\epsilon\,)\mathcal{M}^{(0)} \,,
\nonumber\\
\mathcal{M}^{(2)}(\eps)
&= \mathcal{M}^{(2),\text{fin}}+\bI^{(1)}(\epsilon) \left( \mathcal{M}^{(1),\text{fin}}+\bI^{(1)}(\epsilon\,)\mathcal{M}^{(0)} \right) 
+ \bI^{(2)}(\eps)\mathcal{M}^{(0)} \,.
\label{Mdivcoeffs}
\end{align}
The SCET Wilson coefficient is now obtained by multiplying the IR-divergent amplitude with the inverse of the $\boldsymbol{Z}$-factor. With a slight abuse of notation, we write the expansion of the {\em inverse} $\boldsymbol{Z}$-factor in the form
\begin{align}
\boldsymbol{Z}^{-1}(\epsilon,\,\{p\},\,\mu)&=
1+\frac{\alpha_s}{2\pi} \,\bm{Z}^{(1)}(\eps)
+\left(\frac{\alpha_s}{2\pi}\right)^2 \bm{Z}^{(2)}(\eps)
+\mathcal{O}(\alpha_s^3) \,.
\label{zfac} 
\end{align}
The explicit form of the coefficients $\boldsymbol{Z}^{(n)}(\eps)$ is given in Appendix \ref{renapp}. The above relations can be used to express the $\overline{\text{MS}}$-renormalized amplitude $|\mathcal{M}^{\text{ren}}(\{p\},\mu)\rangle$ in terms of the IR-finite amplitude $|\mathcal{M}^{\text{fin}}(\{p\},\mu)\rangle$
given in \cite{Garland:2002ak,Gehrmann:2011ab,Gehrmann:2013vga,Gehrmann:2011aa}.
At one-loop order, the conversion relation reads
\begin{align}
\mathcal{M}^{(1),\text{ren}}&= 
 \mathcal{M}^{(1),\text{fin}}+ \left(\bI^{(1)}(\epsilon) +\boldsymbol{Z}^{(1)}(\epsilon)\right) \mathcal{M}^{(0)}
\nonumber
\\
&= 
 \mathcal{M}^{(1),\text{fin}}+  \bm{\mathcal{C}_0} \,\mathcal{M}^{(0)}\,,
 \label{eq:oneloopconv}
\end{align}
where $\bm{\mathcal{C}_0} $ is the finite term of Catani's one-loop subtraction operator $\boldsymbol{I}^{(1)}(\epsilon)$,
\begin{align}
\boldsymbol{\mathcal{C}_0}
&=\sum_{(i,j)}\frac{\bT_i\cdot \bT_j}{16} 
\left[\gamma_0^{\text{cusp}} \, \ln^2\frac{\mu^2}{-s_{ij}}
- \frac{4\gamma_0^i}{C_i} \, \ln\frac{\mu^2}{-s_{ij}} \right]
-\frac{\pi^2}{96} \rG'_0\,,
 \label{C0}
\end{align}
with one-loop anomalous dimensions 
$\gamma_0^{\text{cusp}}=4$, $\gamma_0^q=-3C_F$, $\gamma_0^g=-\beta_0$
and $\Gamma_0'=-\gamma_0^{\text{cusp}} \sum_i C_i$.
At two-loop order, the conversion relation takes the form
\begin{align}
\mathcal{M}^{(2),\text{ren}}&= 
\mathcal{M}^{(2),\text{fin}}+
\left(\bI^{(1)}(\epsilon) +\boldsymbol{Z}^{(1)}(\epsilon)\right)
 \mathcal{M}^{(1),\text{fin}} 
\nonumber
\\
&\phantom{=\;}
+ \Big(\bI^{(2)}(\epsilon) + \left(\bI^{(1)}(\epsilon) +\boldsymbol{Z}^{(1)}(\epsilon)\right) \bI^{(1)}(\epsilon) +
\boldsymbol{Z}^{(2)}(\epsilon)\Big)) \mathcal{M}^{(0)}
\nonumber
\\
&= 
\mathcal{M}^{(2),\text{fin}}+
\bm{\mathcal{C}_0} \,  \mathcal{M}^{(1),\text{fin}} 
+\bigg\{ \frac12 \,\bm{\mathcal{C}_0}^2
+\frac{\gamma_1^\text{cusp}}{8}  
\left( \bm{\mathcal{C}_0} +  \frac{\pi^2}{128} \Gamma_0'\right)
\nonumber
\\
&\phantom{=\;}
+ \frac{\beta_0}{2} \left( \bm{\mathcal{C}_1} +  \frac{\pi^2}{32} \bm{\Gamma_0} +  \frac{7\zeta_3}{96} \Gamma_0'  \right)  
-\frac18 \big[ \bm{\Gamma_0} ,\bm{\mathcal{C}_1}  \big]  
\bigg\} \mathcal{M}^{(0)}\,,
\label{eq:twoloopconv}
\end{align}
and the corresponding expression for $\bm{\mathcal{C}_1}$ and the two-loop anomalous dimensions are summarized in Appendix \ref{renapp}. In the appendix, we also give an explicit formula for the commutator $[ \bm{\Gamma_0} ,\bm{\mathcal{C}_1} ]$ in terms of three-particle correlations.  

The above relations are valid for general $n$-parton scattering amplitudes. For $n=3$ colored partons, which is the relevant case here, one can use color conservation to express the dipoles in terms of the Casimir operators associated with the three external legs,
\begin{align}
\bT_1\cdot \bT_2
   &= - \frac12 (C_1 + C_2 - C_3)
   \equiv C_{12}\,,
   \qquad
   \text{etc.}
\end{align}
The color structure then becomes trivial, and the one-loop conversion factor (\ref{C0}) simplifies to
\begin{align}
 \bm{\mathcal{C}_0} 
 &= \frac{C_{12}}{8}\left[ \gamma_0^\text{cusp} 
 \left( \ln^2 \frac{\mu^2}{-s_{12}} - \frac{\pi^2}{6} \right)
 -2 \left( \frac{\gamma_0^1}{C_1} +  \frac{\gamma_0^2}{C_2} \right) \ln \frac{\mu^2}{-s_{12}} \right]
 \nonumber
\\
&\phantom{=\quad}
 + (\text{cyclic permutations})\,.
 \end{align}
The two-loop relation (\ref{eq:twoloopconv}) contains in addition the structure
\begin{align}
&\bm{\mathcal{C}_1} +  \frac{\pi^2}{32} \bm{\Gamma_0} +  \frac{7\zeta_3}{96} \Gamma_0' 
 \nonumber
\\
&\quad
= \frac{C_{12}}{24}\left[ \gamma_0^\text{cusp} 
 \left( \ln^3 \frac{\mu^2}{-s_{12}} 
 + \frac{\pi^2}{4}\ln \frac{\mu^2}{-s_{12}}
 + \frac{3\zeta_3}{2} \right)
 - \left( \frac{\gamma_0^1}{C_1} +  \frac{\gamma_0^2}{C_2} \right) 
 \left( 3\ln^2 \frac{\mu^2}{-s_{12}} + \frac{\pi^2}{4} \right)
 \right]
 \nonumber
\\
&\phantom{\qquad=\;}
 + (\text{cyclic permutations})\,,
\end{align}
as well as a commutator  $[ \bm{\Gamma_0} ,\bm{\mathcal{C}_1}  ]$, which vanishes in the three-parton case because of the trivial color structure.

\section{Hard functions from helicity amplitudes\label{sec:helicity}}

The two-loop four-point helicity amplitudes with one external electroweak boson and three colored particles were computed in \cite{Garland:2002ak,Gehrmann:2011ab,Gehrmann:2013vga,Gehrmann:2011aa}. Having discussed how the amplitudes obtained in these papers can be converted to the $\overline{\text{MS}}$ scheme, we will now show how the hard functions can be assembled from the squared amplitudes. To obtain the result one needs to analytically continue the amplitudes to crossed channels and use parity and charge conjugation symmetry to obtain all helicity configurations from a minimal set. The amplitudes and their analytic continuation to different channels are appended in electronic form to the arXiv submissions of the papers \cite{Garland:2002ak,Gehrmann:2011ab,Gehrmann:2013vga,Gehrmann:2011aa}, since the expressions are too lengthy to be given explicitly in a paper.

\subsection{Kinematics and analytic continuation}

\begin{figure}[t!]
\centering
\psfrag{s12}[lt]{\footnotesize $s_{12}\geq0$}
\psfrag{s13}[lb]{\footnotesize $s_{13}\geq0$}
\psfrag{s23}[b]{\footnotesize $s_{23}\geq0\phantom{abc}$}
\psfrag{1}[]{$\begin{matrix} V\to q \bar{q} g \\ (1) \end{matrix}$}
\psfrag{2}[]{$\begin{matrix} (2) \\  q \bar{q} \to V g  \end{matrix}$}
\psfrag{3}[]{$\begin{matrix} (3) \\  \bar{q} g \to V  \bar{q}  \end{matrix}$}
\psfrag{4}[]{$\begin{matrix} (4) \\  q g \to V  q  \end{matrix}$}
\includegraphics[width=0.6\textwidth]{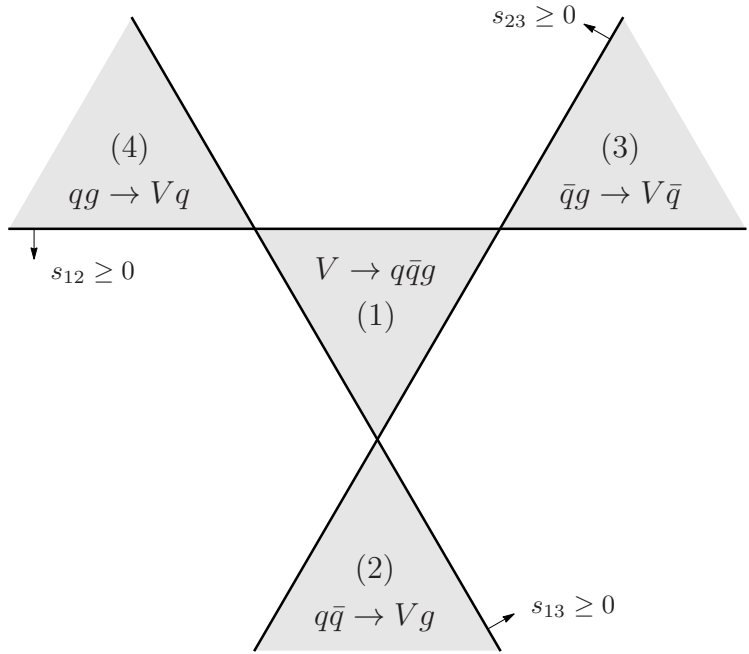}
\caption{Different channels of the $V\to q\bar{q} g$ amplitude. The amplitudes in regions (2), (3) and (4) can be obtained by analytic continuation of the result in region (1). The numbering of regions is the same as in \cite{Gehrmann:2002zr}, where these four regions are denoted by $(1a_+)$, $(2a_+)$, $(3a_+)$, $(4a_+)$.
\label{fig:mandel}}
\end{figure}

We first specify the kinematics and the analytic continuation of the amplitudes, which are common to all channels. For concreteness, we consider the specific process
\begin{align}\label{eq:decay}
 V(q)\rightarrow q(p_1)+\bar{q}(p_2)+g(p_3) \,,
\end{align}
where the vector boson can be off the mass shell. All parton momenta are outgoing and we have
\begin{align}\label{eq:kin}
s_{12}=(p_1+p_2)^2 \,,\quad s_{13}=(p_1+p_3)^2 \,,\quad s_{23}=(p_2+p_3)^2 \,,\quad q^2=(p_1+p_2+p_3)^2\equiv s_{123} \,.
\end{align}
The kinematic region which describes the decay of the electroweak boson to three partons is called region (1) in \cite{Gehrmann:2002zr} and corresponds to the inside of the Mandelstam triangle shown in Figure \ref{fig:mandel}. The amplitudes for vector-boson production can be obtained from the result for $V\to q\bar{q} g$ using crossing symmetry and analytic continuation. The kinematic regions relevant for the considered processes in this paper are (2), (3) and (4) in Figure \ref{fig:mandel}. In the crossed channels, the incoming momenta will enter with a minus sign in the definitions (\ref{eq:kin}). 

The helicity amplitudes for a given process are written in terms of spinor products multiplied by coefficient functions which depend on the invariants $s_{ij}$. In the following, we will denote these coefficient functions by the Greek letters $\alpha_n$, $\beta_n$, $\gamma_n$, $\delta_n$, where the subscript indicates the kinematic region, and use the letter $\Omega$ to denote a generic coefficient function. It is convenient to express the functions in terms of dimensionless invariants. A suitable choice for kinematic region (1) is\footnote{In reference \cite{Gehrmann:2002zr} the notation $y\equiv u_1$ and $z\equiv  v_1$ is used. We use the subscript to distinguish the variables relevant in the different kinematic regions shown in Figure \ref{fig:mandel}.} 
\begin{align}\label{eq:uvdef}
\text{region (1):} \qquad\qquad
u_1 =\frac{s_{13}}{q^2} \,, \qquad\qquad
 v_1=\frac{s_{23}}{q^2}\,.
\end{align}
The coefficient functions are given in \cite{Garland:2002ak,Gehrmann:2011ab,Gehrmann:2013vga,Gehrmann:2011aa} in terms of two-dimensional harmonic polylogarithms (TDHPLs) \cite{Gehrmann:2000zt,Gehrmann:2001ck} in the variables $u_1$ and $v_1$ defined in (\ref{eq:uvdef}). In region (1) these variables fulfill the conditions
\begin{align}\label{kinCond}
0 & < u_1 \leq 1 \,,    & 0 \leq v_1 \leq 1-u_1 \, ,
\end{align}
 and the TDHPLs are analytic and real for these values of the arguments. Since some of the invariants $s_{ij}$ will become negative under crossing, the condition  (\ref{kinCond}) is violated in the regions $(2)$, $(3)$, $(4)$ and one has to analytically continue the amplitudes and the associated TDHPLs. A systematic algorithm for this continuation was given in \cite{Gehrmann:2002zr}. In each region, one defines variables which fulfill conditions analogous to  (\ref{kinCond}) and rewrites the amplitudes in terms of TDHPLs in these variables. They are
\begin{align} 
&\text{region (2):} \qquad\qquad
u_2 = -\frac{s_{13}}{s_{12}} \,, \qquad\qquad
v_2 =\frac{q^2}{s_{12}}\,,
\nonumber\\
&\text{region (3):} \qquad\qquad
u_3 = -\frac{s_{23}}{s_{13}}\,, \qquad\qquad 
v_3 =\frac{q^2}{s_{13}}\,, 
\nonumber\\
&\text{region (4):} \qquad\qquad
u_4 = -\frac{s_{13}}{s_{23}}\,, \qquad\qquad 
v_4 =\frac{q^2}{s_{23}}\,. 
\label{eq:uv}
\end{align} 
To compute the hard functions, we then first take the IR-finite amplitude coefficients in region (i), $\Omega_i^{\text{fin}}(u_i,v_i)$ for $i=2,3,4$ included in the arXiv submissions of \cite{Garland:2002ak,Gehrmann:2011ab,Gehrmann:2013vga,Gehrmann:2011aa}. Using the conversion formulae derived in Section \ref{sec:hard}, these results can then be converted into the $\overline{\text{MS}}$-renormalized coefficicients $\Omega_i^{\text{ren}}(u_i,v_i)$. Note that also  the conversion terms \eqref{eq:oneloopconv} and \eqref{eq:twoloopconv} need to be continued appropriately, but the continuation is trivial, since these terms only contain logarithms of the invariants $s_{ij} \equiv s_{ij} + i0$. To evaluate the TDHPLs numerically, we use the programs \cite{Gehrmann:2001pz,Gehrmann:2001jv}.

\subsection{The \boldmath $q\bar{q}\rightarrow V g$ and $qg\rightarrow Vq$ \unboldmath channels \label{sec:quark}}

The amplitudes for $q\bar{q}\rightarrow V g$ and $qg\rightarrow Vq$ can be obtained from the $V \rightarrow q\bar{q} g$ amplitude using crossing symmetry and analyticity. Note that particles turn into  anti-particles under the crossing operation $p_i\to -p_i$. To obtain the amplitude with a $W^+$ in the final state, one therefore needs to cross the $W^-$ decay amplitude. The hadronic current $\mathcal{S}_{\mu}$ mediating the $V \rightarrow q\bar{q} g $ decay was given in \cite{Garland:2002ak} in the spinor-helicity formalism \cite{Weyl1, Weyl2}. Here, we will use the same spinor notation as \cite{Peskin:2011in} (see e.g.~\cite{Dixon:2011xs,Ellis:2011cr} for overviews of the various types of spinor notations in the literature). The spinor-helicity formalism is a four-dimensional method. The reason it can be applied here is that the $\boldsymbol{Z}$-factor is independent  of the helicities of the partons. The infrared singularities are thus an overall factor in helicity space, which can be removed after which the formalism can be applied to infrared-finite amplitudes. In \cite{Garland:2002ak}, the fixed-helicity current $\mathcal{S}_\mu(q+,\,g+,\,\bar{q}-)$ is written in the form   
\begin{align}\label{eq:helamp}
\mathcal{S}_{\mu}(q+,\,g+,\,\bar{q}-)&=\sqrt{2}R^V_{f_1f_2}\,\bigg\{\alpha_1(u_1,v_1)\;\frac{\langle 1 2\rangle\, [1\,\gamma_\mu\,2\rangle}{\langle 1 3\rangle\,\langle 3 2\rangle}+\beta_1(u_1,v_1)\;\frac{[ 3\,\gamma_\mu\,2\rangle}{\langle 1 3\rangle}\nonumber
\\
 &+\gamma_1(u_1,v_1)\;\frac{[ 3 1] [ 3\,\gamma_\mu\,1\rangle }{\langle1 3 \rangle \,[ 2 3 ]}
+\delta_1(u_1,v_1)\;\frac{[ 3 1 ]}{\langle 1 3\rangle\,[ 2 1]}\left(  [ 1\,\gamma_\mu\,1\rangle +[ 2\,\gamma_\mu\,2\rangle+ [ 3\,\gamma_\mu\,3\rangle\right)\bigg\} \,,
\end{align}
where $R^V_{f_1f_2}$ is the right-handed coupling of the vector boson to the quarks. The electroweak couplings are given explicitly in Table \ref{tab:couplings}. 
\begin{table}[t!]
\begin{center}
\begin{tabular}{lcccc}\toprule
	$V$ & $\gamma$ & $Z$ & $W^+$ & $W^-$
	\\
	\midrule
	$L_{ij}^V$ & $e_i \delta_{ij}$
	& $\frac{I^3_{i}-\sin^2(\theta_W)e_i}{\sin{\theta_W}\cos{\theta_W}} \delta_{ij}$
	& $\frac{1}{\sqrt{2}\sin{\theta_W}} I^+_{ij} V_{ij}^*$
	& $\frac{1}{\sqrt{2}\sin{\theta_W}} I^-_{ij} V_{ji}$
	\\
	$R_{ij}^V$ & $e_i \delta_{ij}$
	& $\frac{-\sin{\theta_W}e_i}{\cos{\theta_W}} \delta_{ij}$
	& $0$
	& $0$
	\\
	\bottomrule
\end{tabular}
\end{center}
\caption{Couplings of quarks with flavors $i$ and $j$ to electroweak vector bosons in the final state. The charges of up-type and down-type quarks are $e_{u} = \frac{2}{3}$, $e_{d} = -\frac{1}{3}$, electroweak isospin $I^3_{u} = \frac{1}{2}$, $I^3_{d} = -\frac{1}{2}$, $I^+_{ud} = I^-_{du} =1$, $I^\pm_{uu} = I^\pm_{dd} =0$. $V_{ij}$ is the CKM matrix.\label{tab:couplings}}
\end{table}
The coefficients $\alpha_1$, $\beta_1$, $\gamma_1$, $\delta_1$ are expanded as
\begin{equation}\label{eq:qqVgOmega}
\Omega= e\, g_s\,(t^a)_{ij}\,\left(\Omega^{(0)}+\frac{\alpha_s}{2\pi}\,\Omega^{(1)}+\left(\frac{\alpha_s}{2\pi}\right)^2\Omega^{(2)}+ \dots \right) \,,
\end{equation}
where $i$, $j$ and $a$ are the color indices of the quark, anti-quark and gluon. We note that the coefficient $\delta_1(u_1,v_1)$ is not independent, as it is linked to the other functions by current conservation. To obtain the SCET hard functions, we need to evaluate the amplitudes with the renormalized coefficients $\Omega^{\rm ren}$ in the $\overline{\text{MS}}$ scheme. The conversion from the IR-finite amplitudes of \cite{Garland:2002ak,Gehrmann:2011ab} to the $\overline{\text{MS}}$ scheme was discussed in detail in the previous section. Since the IR singularities are independent of the spin of the particles, the coefficients are converted using the same expressions \eqref{eq:oneloopconv} and \eqref{eq:twoloopconv}. We will suppress the superscript $\Omega^{\text{ren}}$ on the amplitude coefficients in the following, but it is understood that the renormalized quantities are used to obtain the hard functions.

The remaining helicity configurations follow from \eqref{eq:helamp} by using parity conservation and charge conjugation symmetry of the strong interaction. Parity yields the relation
\begin{align}\label{eq:parity}
\mathcal{S}_{\mu}(q-,\,g\lambda,\,\bar{q}+)=\mathcal{S}^*_{\mu}(q+,\,g(-\lambda),\,\bar{q}-)   \bigg|_{R^V_{f_1f_2} \to L^V_{f_1f_2}} \,,
\end{align}
and charge conjugation
\begin{align}\label{eq:charge}
\mathcal{S}_{\mu}(q\lambda_q,\,g\lambda,\,\bar{q}\lambda_{\bar{q}})=(-1)\,\mathcal{S}_{\mu}(\bar{q}\lambda_{\bar{q}},\,g\lambda,\,q\lambda_q) \,.
\end{align}

To obtain the hard functions in a given channel, we need to square the helicity amplitudes. Summing the contribution of different helicities, we then get the hadronic tensor
\begin{align}
H_{\mu\nu}=\sum_{\text{helicities}} \mathcal{S}_{\mu}\,\mathcal{S}^{*}_{\nu} \,.
\end{align}
By contracting this with the appropriate lepton tensor, one could obtain the vector-boson production cross section with an arbitrary set of cuts on the leptons arising in the vector-boson decay. Here, we will only be interested in the total vector-boson production rate and we thus only need the hard function 
\begin{equation}
H = -g^{\mu\nu}\,H_{\mu\nu} \,.
\end{equation}
The contribution of the helicity configuration $\mathcal{S}_\mu(q+,\,g+,\,\bar{q}-)$ to the hard function in the process $V\to q\bar{q} g$, for example, is obtained from the representation (\ref{eq:helamp}) and has the form
\begin{equation}\label{onehel}
- \mathcal{S}^*_\mu(q+,\,g+,\,\bar{q}-) \mathcal{S}^\mu(q+,\,g+,\,\bar{q}-) = 2|R^V_{f_1f_2}|^2 \, h_1(s_{12},s_{23},s_{13}) \,.
\end{equation}
The subscript on the quantity $h_1$ denotes the region in which the amplitudes are evaluated. In region $(n)$, we obtain
\begin{align}
h_{n}(s_{12},s_{23},s_{13})=\frac{1}{s_{13}\,s_{23}\,q^2}&\bigg\{
|\alpha_n(u_{n},v_{n})|^2\,s_{12}\left(2 s_{12} q^2+s_{13}s_{23}  \right)\nonumber\\
&+|\beta_n(u_n,v_n)|^2\,s_{23}\left(2 s_{23} q^2+s_{12} s_{13}    \right)\nonumber\\
&+|\gamma_n(u_n,v_n)|^2\,s_{13}\left( 2 s_{13} q^2+s_{12} s_{23}  \right)\nonumber\\
&+2\Re\left[\alpha_n(u_n,v_n)\,\beta_n^*(u_n,v_n)\right]\,s_{12}s_{23}\left( 2 q^2-s_{13}\right)\nonumber\\
&-2\Re\left[\alpha_n(u_n,v_n)\,\gamma_n^*(u_n,v_n)\right]\,s_{12} s_{13} s_{23}\nonumber\\
&-2\Re\left[\beta_n(u_n,v_n)\,\gamma_n^*(u_n,v_n)\right]\,s_{13}s_{23}\left( 2 q^2-s_{12} \right)\bigg\}\,.
\end{align}
It is understood that the variables $u_{n}\equiv u_{n}(s_{12},s_{23},s_{13}) $ and $v_{n}\equiv v_{n}(s_{12},s_{23},s_{13}) $ in the amplitude coefficients are expressed in the invariants $s_{ij}$ via the relations (\ref{eq:uv}) valid in region $(n)$. 

The full hard function is obtained by summing over all helicities. Since the other helicity configurations can be obtained by applying parity and charge conjugation to $\mathcal{S}_\mu(q+,\,g+,\,\bar{q}-)$, see (\ref{eq:parity}) and (\ref{eq:charge}), the result can be expressed using the same function $h_1$. For $V\to q\bar{q} g$, one then finds
\begin{equation}\label{HadTens1a}
H_{V\to q\bar{q} g}=  2\left(|R^V_{f_1f_2}|^2+|L^V_{f_1f_2}|^2\right) \left[ h_1(s_{12},s_{23},s_{13})+ h_1(s_{12},s_{13},s_{23})\right]\,.
\end{equation}
The contribution with $s_{13} \leftrightarrow s_{23}$ arises from charge conjugation. For the channel $q\bar{q}\rightarrow V g$, one obtains the same expression, but the amplitudes now have to be continued to region (2) and evaluated with the variables $u_2$ and $v_2$,
\begin{equation}\label{eq:reg2}
H_{q\bar{q}  \to V g} = 2\left(|R^V_{f_1f_2}|^2+|L^V_{f_1f_2}|^2\right) \left[ h_2(s_{12},s_{23},s_{13})+ h_2(s_{12},s_{13},s_{23})\right]\,.
\end{equation}
Finally, for region (4), $q g\rightarrow V q$, we need
\begin{equation}\label{eq:reg4}
H_{qg  \to V q } = - 2\left(|R^V_{f_1f_2}|^2+|L^V_{f_1f_2}|^2\right) \left[ h_4(s_{12},s_{23},s_{13})+ h_3(s_{12},s_{13},s_{23})\right]\,.
\end{equation}
The extra minus sign in front of \eqref{eq:reg4} compensates the minus sign which arises when crossing a fermion from the final to the initial state. The other difference to the previous two cases is that charge conjugation maps region $(4)$ onto region $(3)$, while $(1)$ and $(2)$ map onto themselves. For this reason, the result \eqref{eq:reg4} involves the amplitude in both regions. 
Note that the variables $s_{ij}\equiv2\sigma_{ij} p_i\cdot p_j$ are the same in all kinematic regions. They stay invariant under crossing because of the sign factor which is 
$\sigma_{ij}=+1$ if the momenta $p_i$ and $p_j$ are both incoming or outgoing, and $\sigma_{ij}=-1$ otherwise. For $q\bar{q}\rightarrow Vg$ they relate to the usual partonic Mandelstam variables in equation (\ref{eq:factorization}) via $\hat{s}=s_{12}$, $\hat{t}=s_{13}$ and $\hat{u}=s_{23}$, while the relations are $\hat{s}=s_{23}$, $\hat{t}=s_{13}$ and $\hat{u}=s_{12}$ for $q g\rightarrow Vq$. The hard function for $H_{gq  \to V q }$ is obtained from $H_{qg  \to V q }$ by exchanging $\hat{t}$ and $\hat{u}$. The hard functions for anti-quark channels are equal to the ones for the quarks, $H_{\bar{q}g  \to V \bar{q} }=H_{qg  \to V q }$ and $H_{g\bar{q}  \to V \bar{q} }=H_{g q  \to V q }$.

\subsection{\boldmath The $gg \to V g$ channel \unboldmath\label{sec:gluon}}

In Section \ref{sec:hard} the SCET operators for the partonic channel $gg \rightarrow V g$ were given and we now extract their Wilson coefficients from the helicity amplitudes provided in \cite{Gehrmann:2013vga}. Since we only need the leading-order amplitude, which is free of infrared divergences, we can directly use the result presented in \cite{Gehrmann:2013vga}. In analogy to \cite{Garland:2002ak}, the helicity amplitudes are first given in region $(1)$, corresponding to $V\to g g g$ and then analytically continued into the other regions. However, instead of the hadronic current, \cite{Gehrmann:2013vga} provides the result after contraction with the lepton tensor
\begin{align}
l^-(p_5)+l^+(p_6)\rightarrow V(q)\rightarrow g(p_1)+g(p_2)+g(p_3)\,,
\end{align}
which for a right-handed lepton has the form
\begin{equation}
L_R^{\mu}(p_5^+,\,p_6^-)=[6\gamma^\mu 5\rangle\,.
\end{equation}
This contraction then gives a set of helicity amplitudes. Denoting the vector part of the coupling of the boson $V$ to the quarks inside the loop by
\begin{align}
Q_V^g = \frac{1}{2} \sum_q (L_{qq}^V + R_{qq}^V) \,,
\end{align}
the amplitudes for the configurations $(p_1^+,\,p_2^-,\,p_3^-)$ and $(p_1^+,\,p_2^+,\,p_3^+)$  have the form
\begin{align}
A_R^{(+--)}&(p_5,p_6;p_1,p_2,p_3) = L_R^\mu(p_5^+,p_6^-) \, S_\mu(p_1^+,p_2^-,p_3^-) = 
\frac{e \,Q_V^g}{\sqrt{2}} \, \frac{\langle 2\,3\rangle }{\langle 1\, 2\rangle \langle 1\, 3\rangle  [2\,3]} \nonumber \\ \times  \Bigg\{ 
& \langle 2\,5\rangle \langle 3\,5\rangle  [5\,6] \, \alpha^a_n(u_n,v_n) 
+ \langle 2\,3\rangle \langle 2\,5\rangle  [2\,6] \, \alpha^b_n(u_n,v_n)
+ \langle 2\,3\rangle \langle 3\,5\rangle  [3\,6] \, \alpha^c_n(u_n,v_n) \, \Bigg\}\,, \label{apmm}
\end{align}
and
\begin{align}
 A_R^{(+++)}&(p_5,p_6;p_1,p_2,p_3) = L_R^\mu(p_5^+,p_6^-) \, S_\mu(p_1^+,p_2^+,p_3^+) =
\frac{e \,Q_V^g}{\sqrt{2}} \nonumber \\\times  \Bigg\{ 
&\frac{[1\,3] \langle 1\,5\rangle  [1\,6]}{\langle 1\, 2\rangle \langle 2\,3\rangle} \beta^a_n(u_n,v_n)
+\frac{[2\,3] \langle 2\,5\rangle  [2\,6]}{\langle 1\, 2\rangle \langle 1\,3\rangle} \beta^b_n(u_n,v_n)
+\frac{[2\,3] \langle 2\,5\rangle  [1\,6]}{\langle 1\, 2\rangle \langle 2\,3\rangle} \beta^c_n(u_n,v_n)    \Bigg\}\,, \label{appp}
\end{align}
where the coefficients $\alpha_n^i$  and $\beta_n^ i$ can be found in the electronic appendix to the arXiv submission of \cite{Gehrmann:2013vga}.\footnote{In \cite{Gehrmann:2013vga} the coefficients $\alpha_n^a$, $\alpha_n^b$ and $\alpha_n^c$ are denoted by $\alpha_{1}$, $\alpha_{2}$ and $\alpha_{3}$, and analogously for the $\beta$-coefficients. As before, we use the subscript to denote the kinematic region. In addition \cite{Gehrmann:2013vga} does not include the prefactor $e \,Q_V^g$ in the definition of the amplitudes $A_R$. } Their expansion is written in the form
\begin{equation}
\Omega=g_s\,d^{abc}\,\left(\frac{\alpha_s}{2\pi}\,\Omega^{(1)}+\left(\frac{\alpha_s}{2\pi}\right)^2\Omega^{(2)}+ \dots \right) \,.
\end{equation}
The color factor of the hard function, obtained from the squared amplitudes, is $(d^{abc})^2 = 40/3$. Let us note that there is also an axial-vector contribution for the case $V=Z$ at leading order \cite{vanderBij:1988ac,Hopker:1993pb}, which is not given in \cite{Gehrmann:2013vga} and will not be included in the following. All other helicity amplitudes follow from \eqref{apmm}, \eqref{appp} by permutation of the external legs, parity conjugation, and the relation $A_L^{(i)}(p_5,p_6;p_1,p_2,p_3)=A_R^{(i)}(p_6,p_5;p_1,p_2,p_3)$, see \cite{Gehrmann:2013vga}.

To obtain the hard function for vector-boson production, we first square the amplitudes and then integrate over the angle of the leptons in the rest frame of the vector boson to remove the lepton tensor. 
The integral of the lepton tensor over the direction of the leptons takes the form
\begin{align}
 \int\frac{d\Omega}{4\pi}\, L_R^{\mu}(p_5^+,\,p_6^-) {L_R^{\nu}}^*(p_5^+,\,p_6^-)&=\int\frac{d\Omega}{4\pi}\, [6\gamma^\mu 5\rangle\,[5\gamma^\nu 6\rangle  
=-\frac{2}{3}q^2\left(g^{\mu\nu}-\frac{q^\mu q^\nu}{M_V^2}\right)\,.
\end{align}
The same result is obtained for the left-handed current $L_L^{\mu}(p_5^-,\,p_6^+)= L_R^{\mu}(p_6^+,\,p_5^-)$. Due to current conservation, the contraction of $q_\mu$ with the hadron tensor vanishes so that, up to a prefactor, the integration over the direction is the same as contracting with $g_{\mu\nu}$. The contribution of the right-handed lepton to the hard function in region $(1)$ for the helicity configuration $(+--)$, for example, is obtained from
\begin{align}
h_1^{(+--)}(s_{12},s_{23},s_{13})= -S^*_\mu(p_1^+,p_2^-,p_3^-) S^\mu(p_1^+,p_2^-,p_3^-)=\frac{3}{2q^2}\int\frac{d\Omega}{4\pi}\,  \Big|A_R^{(+--)}&(p_5,p_6;p_1,p_2,p_3)  \Big|^2\,,
\end{align}
and the full hard function is obtained by summing over the helicities. To perform the angle integrations, we first rewrite the product of spinor products in terms of traces, using identities such as
\begin{align}
[i,\,j]\langle j,\,k\rangle [k,\,l]\langle l,\,i\rangle = \text{Tr}\left[ \frac{1+\gamma_5}{2}\,\, \dsl{p}_i \dsl{p}_j \dsl{p}_k \dsl{p}_l \right]\,.
\end{align}
Since the result after the angle average only depends on the three gluon momenta, the $\gamma_5$ term in the trace does not contribute. We then obtain the following results for the two helicity configurations $h_n^{(+--)}$ and $h_n^{(+++)}$ expressed in terms of the coefficients $\alpha^{i}_n$ and $\beta^{i}_n$ defined in equations (\ref{apmm}) and (\ref{appp}),
\begin{align}
 h_n^{(+--)}(s_{12},s_{23}, s_{13}) &=\frac{(e \,Q_V^g)^2}{8q^2\, s_{12} s_{13}}\bigg\{\nonumber\\
&\phantom{+}\;\,|\alpha^a_n(u_n,v_n)|^2\,q^2\,(s_{23}\,q^2+2 s_{12}\, s_{13})
\nonumber\\
&+|\alpha^b_n(u_n,v_n)|^2\,s_{23}\,(s_{12}+s_{23})^2
+|\alpha^c_n(u_n,v_n)|^2\,s_{23}\,(s_{13}+s_{23})^2
\nonumber\\
&-2\Re\left[ \alpha^a_n(u_n,v_n)\alpha^{b*}_n(u_n,v_n) \right]\,s_{23}\,(s_{12}+s_{23})\,q^2\nonumber\\
&+2\Re\left[ \alpha^a_n(u_n,v_n)\alpha^{c*}_n(u_n,v_n) \right]\,s_{23}\,(s_{13}+s_{23})\,q^2\nonumber\\
&-2\Re\left[ \alpha^b_n(u_n,v_n)\alpha^{c*}_n(u_n,v_n) \right]\,s_{23}\left(s_{23}\, q^2- s_{12} s_{13}  \right)\bigg\}
\end{align}
and
\begin{align}
 h_n^{(+++)}(s_{12},s_{23}, s_{13})&=\frac{(e \,Q_V^g)^2}{8q^2\, s_{12} }\bigg\{
\nonumber\\
&\phantom{+}\;\,|\beta^a_n(u_n,v_n)|^2\,\frac{s_{13}\,(s_{12}+s_{13})^2}{s_{23}}
+|\beta^b_n(u_n,v_n)|^2\,\frac{s_{23}\,(s_{12}+s_{23})^2}{s_{13}}
\nonumber\\
&+|\beta^c_n(u_n,v_n)|^2\,\left( 2\,s_{12}\,q^2 +s_{13}\,s_{23} \right)
\nonumber\\
&-2\Re\left[ \beta^a_n(u_n,v_n)\beta^{b*}_n(u_n,v_n) \right]\,\left( s_{12} \,q^2-s_{13}\,s_{23} \right)
\nonumber\\
&+2\Re\left[\beta^a_n(u_n,v_n)\beta^{c*}_n(u_n,v_n) \right]\,s_{13}\,(s_{12}+s_{13})
\nonumber\\
&+2\Re\left[ \beta^b_n(u_n,v_n)\beta^{c*}_n(u_n,v_n) \right]\,s_{23}\,(s_{12}+s_{23})\bigg\} \,.
\end{align}
As before, the subscript on $h_n^{(i)}$ indicates that this is the result in the kinematic region $(n)$. 

Opposite helicity configurations give identical contributions, and the remaining configurations are obtained by permuting the gluon momenta. Using these relations, the full hard function in region (1) is obtained as
\begin{align}\label{hardggOne}
H_{V \to ggg} =-g_{\mu\nu} H^{\mu\nu} 
& =2\left[  h_1^{(+++)}(s_{12},s_{23}, s_{13}) + h_1^{(+--)}(s_{12},s_{23}, s_{13}) \right. \nonumber\\ 
&\hspace{2cm}  \left. +  h_1^{(-+-)}(s_{12},s_{23}, s_{13})+ h_1^{(--+)}(s_{12},s_{23}, s_{13}) \right]\,\nonumber \\
&= 2\left[ h_1^{(+++)}(s_{12},s_{23}, s_{13}) + h_1^{(+--)}(s_{12},s_{23}, s_{13}) \right.\nonumber\\
 &\hspace{2cm} \left.+h_1^{(+--)}(s_{12},s_{13}, s_{23})+h_1^{(+--)}(s_{23},s_{12}, s_{13})\right]\,.
\end{align}
The factor 2 accounts for the opposite helicity contributions.

We are interested in the result in the kinematic region $(2)$, which corresponds to the process $g(p_1) + g(p_2) \to g(p_3) + V(q)$. Proceeding exactly as discussed in the previous subsection, one first analytically continues the amplitudes $\alpha^{i}_1$ and $\beta^{i}_1$ to region (2) and expresses them in the kinematic variables $u_2$ and $v_2$ relevant in this region. The last term in equation (\ref{hardggOne}) requires to exchange $p_1$ and $p_3$, and one therefore also needs the amplitude in region (4),  which describes the process $g(p_2) + g(p_3) \to g(p_1) + V(q)$. The result for the full hard function in region $(2)$ then reads
\begin{align}\label{hardggTwo}
H_{gg\to Vg} = -g_{\mu\nu} H^{\mu\nu} 
=& 2\left[ h_2^{(+++)}(s_{12},s_{23}, s_{13}) + h_2^{(+--)}(s_{12},s_{23}, s_{13}) \right.\nonumber\\ 
&\hspace{2cm}\left. +h_2^{(+--)}(s_{12},s_{13}, s_{23})+h_4^{(+--)}(s_{23},s_{12}, s_{13})\right]\,.
\end{align}
The analytically continued amplitudes are  included  in the arXiv submission of \cite{Gehrmann:2013vga}. 

To get the lowest-order cross section in the $gg\to Vg$ channel, we need to average over the spins and colors of the incoming gluons. This gives
\begin{equation}
 \hat{s} \frac{\rd\hat{\sigma} }{\rd\hat{u}\, \rd\hat{t}} 
 =\frac{1}{16\pi \hat{s}}\,\frac{1}{256} \,H_{gg\to Vg}(\hat{u},\hat{t},\mu) \,\delta(m_X^2) 
\, \equiv\, \hat{\sigma}_{gg}^{(0)}(\hat{u},\hat{t},\mu) \,\hat{H}_{gg\to Vg}(\hat{u},\hat{t},\mu) \,\delta(m_X^2)  \,.
\end{equation}
In the resummed cross section, the hard function will be multiplied by a convolution of the gluon jet function $J_g$ with the soft function $S_{gg}$, see (\ref{eq:factorization}). To write the factorization theorem in the form  (\ref{eq:factorization}), we have introduced a hard function $\hat{H}_{gg\to Vg}$, which is normalized to one at lowest order. However, the leading-order cross section $\hat{\sigma}_{gg}^{(0)}$ is of  $\mathcal{O}(\alpha_s^3)$ instead of $\mathcal{O}(\alpha_s)$ as in the other channels.

\subsection{Hard functions for Higgs production\label{sec:higgs}}

The main focus of our paper is on vector-boson production, but for completeness we now also provide the hard functions for Higgs production in association with a jet, since the construction is completely analogous to the vector-boson case. The factorization theorem \eqref{eq:factorization} is valid also for Higgs production and it involves the same jet and soft functions as in the vector-boson case. With the hard functions given here, the resummation can be extended to N$^3$LL accuracy also for Higgs production. 

Since the infrared singularities of the amplitudes are independent of the spin, the conversion from Catani's subtraction scheme to $\overline{\text{MS}}$ is obtained with exactly the same formulae \eqref{eq:oneloopconv} and \eqref{eq:twoloopconv} as in the vector-boson case. The two-loop helicity amplitudes for Higgs production in the heavy top-quark limit were given in \cite{Gehrmann:2011aa}. For $H\to g(p_1) + g(p_2) + g(p_3)$, one has
\begin{align}
\mathcal{M}_{ggg}^{+++}(p_1,p_2, p_3) &= \frac{1}{\sqrt{2}}\frac{M_H^4}{\langle 12 \rangle \langle 23 \rangle \langle 31 \rangle} \, \alpha_1(u_1,v_1) \,, \nonumber \\
\mathcal{M}_{ggg}^{++-}(p_1,p_2, p_3) &=  \frac{1}{\sqrt{2}}\frac{[1 2]^3}{[ 23 ] \, [1 3]}\,\beta_1(u_1,v_1)   \,.
\end{align}
The dimensionless variables $u_n$ and $v_n$  relevant in the different kinematic regions were defined in (\ref{eq:uvdef}) and  (\ref{eq:uv}) and the expansion of the coefficient functions $\alpha_n,\beta_n$ is now written as
\begin{equation}\label{eq:higgsOmega}
\Omega=\frac{\alpha_s C_t}{3\pi v}\, g_s\,f^{abc}\,\left(\Omega^{(0)}+\frac{\alpha_s}{2\pi}\,\Omega^{(1)}+\left(\frac{\alpha_s}{2\pi}\right)^2\Omega^{(2)}+ \dots \right) \,,
\end{equation}
where $C_t$ is the Wilson coefficient of the effective Lagrangian
\begin{equation}\label{eq:Leff}
{\cal L}_{\rm eff} =  C_t\,\frac{\alpha_s}{12\pi} \frac{H}{v}\, G_{\mu\nu}^a G^{\mu\nu,a}\,,
\end{equation}
which mediates Higgs production in the large $m_t$-limit and $v$ is the vacuum expectation value of the Higgs field. The two-loop value of $C_t$ is listed in Appendix \ref{app:jetsoft}.
When squaring the amplitudes, the color factor is $(f^{abc})^2 = C_A\, d_A$, where $d_A=N_c^2-1$ is the dimension of the adjoint representation. We obtain
\begin{align}
h_n^{(+++)}(s_{12},s_{23},s_{13})&= 
\left | \mathcal{M}_{ggg}^{+++}(p_1,p_2, p_3) \right|^2   = \frac{M_H^8}{2 s_{12} s_{23} s_{13}} |\alpha_n(u_n,v_n)|^2 \, ,\\
h_n^{(++-)}(s_{12},s_{23},s_{13}) &= \left | \mathcal{M}_{ggg}^{++-}(p_1,p_2, p_3) \right|^2=  \frac{s_{12}^3}{2 s_{23} s_{13}} |\beta_n(u_n,v_n)|^2\,.
\end{align}
The remaining amplitudes follow by using parity conservation and symmetry under the exchange of the final state gluons. Summing over all helicities, the hard function for Higgs production in the gluonic channel $gg\to Hg$ becomes
\begin{align}
H_{gg\to H g} &= 2\, \left[ h_2^{(+++)}(s_{12},s_{23},s_{13})  
+  h_2^{(++-)}(s_{12},s_{23},s_{13}) \right. \nonumber \\ & \left.\hspace{2cm}
+  h_4^{(++-)}(s_{23},s_{12},s_{13})+ h_4^{(++-)}(s_{13},s_{12},s_{23}) \right]\,.
\end{align}
The factor of two accounts for the equal opposite-helicity contributions. This result looks different than the hard function for $gg \to V g$ in \eqref{hardggTwo} because the corresponding hard function was obtained from the helicity configuration $(+--)$ instead of $(++-)$. The one-loop hard function $H_{gg\to H g} $ was also given in \cite{Jouttenus:2013hs}; we agree with this result.

For the quark channel, $H\to q(p_1) + \bar{q}(p_2) + g(p_3)$, there is a single independent amplitude,
\begin{align}
\mathcal{M}_{q\bar{q}g}^{-++}(p_1,p_2, p_3) &= \frac{1}{\sqrt{2}}\frac{[2 3]^2}{[ 1 2 ]}\, \gamma_1(u_1,v_1) \,.
\end{align}
The expansion of the coefficient $\gamma_n$ is written in the form \eqref{eq:higgsOmega}, but with color structure $(t^a)_{ij}$ instead of $f^{abc}$. This yields a factor $(t^a)_{ij} (t^a)_{ji}=C_F\,d_F$ (where $d_F=N_c$ is the dimension of the fundamental representation) in the squared amplitude, which has the form 
\begin{equation}
h_n(s_{12},s_{23},s_{13}) =\left|\mathcal{M}_{q\bar{q}g}^{-++}(p_1,p_2, p_3)\right|^2 = \frac{s_{23}^2}{2 s_{12}}\, |\gamma_n(u_n,v_n)|^2\,.
\end{equation}
Using parity and charge conjugation, one derives the other helicity configurations. For the full hard functions, we then obtain, in analogy to (\ref{eq:reg2}) and (\ref{eq:reg4}),
\begin{align}
H_{q\bar{q}\to H g} &= 2 \left[ h_2(s_{12},s_{23},s_{13})+ h_2(s_{12},s_{13},s_{23})\right] \,,\\
H_{q g\to H g } &= -2 \left[ h_4(s_{12},s_{23},s_{13})+ h_3(s_{12},s_{13},s_{23})\right]\,,
\end{align}
where the factor 2 accounts for the identical, parity-opposite contributions, and the crossed terms arise from charge conjugation. The coefficients $\alpha_n$, $\beta_n$ and $\gamma_n$ can be obtained in electronic form from the source files of the arXiv version of \cite{Gehrmann:2011aa}.

\section{Two-loop cross section near threshold\label{sec:expansion}}
The singular threshold cross section can be obtained by performing the fixed-order expansion of the resummed partonic cross sections, whose explicit expressions are given in \cite{Becher:2009th,Becher:2012xr}. In these papers, the resummation is achieved by solving renormalization group (RG) equations for the hard, jet and soft functions and evolving them to the factorization scale $\mu_f$, where they can then be combined with the PDFs. This allows one to evaluate each contribution at its characteristic scale. For example, to avoid large logarithms in the hard function, the starting scale $\mu_h$ of the RG evolution is chosen to be $\mu_h\approx p_T$. For the jet and soft functions lower starting scales $\mu_j$ and $\mu_s$ are appropriate, as discussed in detail in \cite{Becher:2009th,Becher:2012xr}. 

The simplest way of going back to the fixed-order expressions is to switch off the resummation by taking the limit where the scales $\mu_h$, $\mu_j$, $\mu_s$ and $\mu_f$ all become equal. In Laplace space, where the cross section factors into a product of the hard, jet and soft functions, taking the limit is completely trivial. All the RG-evolution factors switch off and the fixed-order result is simply the product of the Laplace-transformed functions. In momentum space, the limit is a bit more delicate since the cross section becomes distribution valued in this limit. Starting with (20) in \cite{Becher:2012xr} and taking the limit in which all scales coincide whenever it is trivial, we are left with
\begin{align}
\frac{d^2\sip_{abc}^{\text{sing}}}{dy\,dp_T^2}&=\lim_{\eta\rightarrow 0}\;\sip_{ab}^{(0)}(\up,\,\tp,\,\mu)\,{\hat H}_{ab}(\up,\,\tp,\,\mu) \tilde{j}_{c}(\partial_{\eta},\,\mu)
 \tilde{s}_{ab}\left( \partial_{\eta}+\ln\,\frac{\mu}{p_T},\,\mu \right)\,\frac{1}{m_X^2}\left(\frac{m_X^2}{\mu^2} \right)^{\eta}\,\frac{e^{-\gamma_E\eta}}{\Gamma(\eta)} \,,
\end{align}
for the channel $a+b \to V+ j_c $. To factor out the tree-level cross section $\sip_{ab}^{(0)}$, we have normalized the hard functions ${\hat H}_{ab}$  to one at the lowest order and have indicated this by putting a hat on the normalized functions. The jet function $\tilde{j}_{c}$ and the soft function $\tilde{s}_{ab}$ appearing here are the Laplace-transformed functions. Their explicit form can be found in Appendix \ref{app:jetsoft}. At the $n$-th order in perturbation theory, the functions are polynomials of order $2n$ in logarithms of the Laplace variables. In the above representation, these logarithms are replaced by derivatives acting on an $m_X$-dependent kernel. This type of solutions for the RG equations of the jet and soft functions was introduced in \cite{Becher:2006nr}. To take the limit $\eta \to 0$, we first need to expand this kernel in a series of distributions
\begin{align}\label{eq:star}
\frac{1}{m_X^2}\left(\frac{m_X^2}{\mu^2} \right)^{\eta}&= \frac{1}{\eta}\,\delta(m_X^2)   +\sum_{n=0}^\infty\,\frac{\eta^n}{n!}\left[ \frac{\ln^n\frac{m_X^2}{\mu^2}}{m_X^2} \right]_\star \,.
\end{align}
The $\star$-distributions appearing on the right-hand side are generalizations of the usual plus-distributions to dimensionful variables. Their explicit form can be obtained by rewriting the integration over the invariant mass $m_X$ in the form
\begin{align}
\int_0^{m^2_\text{max}}dm_X^2\,\frac{1}{m_X^2}&\left(\frac{m_X^2}{\mu^2} \right)^{\eta}\,f(m_X^2)\nonumber\\
&=\frac{1}{\eta} \left(\frac{m_\text{max}^2}{\mu^2} \right)^{\eta}\,f(0)
+ \int_0^{m^2_\text{max}}dm_X^2\,\frac{1}{m_X^2}\left(\frac{m_X^2}{\mu^2} \right)^{\eta}\,\left[ f(m_X^2)-f(0)  \right]
\end{align}
and expanding the right-hand side in powers of $\eta$, which yields
\begin{align}
\int_0^{m^2_\text{max}}dm_X^2\,\left[ \frac{\ln^n\frac{m_X^2}{\mu^2}}{m_X^2} \right]_\star\,f(m_X^2)
= \frac{f(0)}{n+1} \ln^{n+1}\frac{m^2_\text{max}}{\mu^2}+\int_0^{m^2_\text{max}}dm_X^2\,\frac{\ln^n\frac{m_X^2}{\mu^2}}{m_X^2}\left[ f(m_X^2)-f(0)  \right]\,.
\end{align}

The expansion of the cross section is now straightforward. To present the result, we write the perturbative expansion of the normalized hard function in the form
\begin{equation}
\hat{H}_{ab}(\hat{u},\hat{t},\mu) =1+\sum_{n=1}^\infty \left(\frac{\alpha_s}{4\pi}\right)^n \,h^{(n)}\,
\end{equation}
and also introduce expansion coefficients $p^{(n)}_{i}$ which capture the contribution of the product of the jet and soft function at the $n$-th order in perturbation theory. The coefficients  $p^{(n)}_{0}$ multiply $\delta(m_X^2)$ and the higher coefficients $0<i\leq 2n$ the $\star$-distributions arising in the expansion (\ref{eq:star}). The coefficients $h^{(n)}$ and $p^{(n)}_{i}$ depend on the partonic channel, but in the following we suppress the channel indices $a$, $b$ and $c$ for better readability. To two-loop order, the cross section then has the structure
\begin{multline}\label{sigmaexpanded}
\frac{d^2\sip^{\text{sing}}}{dy\,dp_T^2}=\sip^{(0)}\Bigg\{\delta(m_X^2)+\frac{\alpha_s}{4\pi}\Bigg[\delta(m_X^2)\left( p^{(1)}_{0}+h^{(1)} \right)+\left[ \frac{1}{m_X^2} \right]_\star p^{(1)}_{1}+\left[ \frac{\ln\frac{m_X^2}{\mu^2}}{m_X^2} \right]_\star p^{(1)}_{2}\Bigg]\\
+\left( \frac{\alpha_s}{4\pi}\right)^2\Bigg[ \delta(m_X^2)\left( h^{(2)}+h^{(1)}\cdot p_0^{(1)}+p_0^{(2)} \right)
   +\left[ \frac{1}{m_X^2} \right]_\star\left( h^{(1)}\cdot p_1^{(1)}+p_1^{(2)} \right) \\
  +\left[ \frac{\ln\frac{m_X^2}{\mu^2}}{m_X^2} \right]_\star\left( h^{(1)}\cdot p_2^{(1)}+p_2^{(2)} \right)   +\left[ \frac{\ln^2\frac{m_X^2}{\mu^2}}{m_X^2} \right]_\star p_3^{(2)}+\left[ \frac{\ln^3\frac{m_X^2}{\mu^2}}{m_X^2} \right]_\star p_4^{(2)}\Bigg]\Bigg\}\,.
\end{multline}
The explicit form of the one-loop coefficients in the above formula is
\begin{align}
p_0^{(1)}&=-\frac{\pi^2\,\gamma_0^\text{cusp}}{12}\left(C_J+4C_S\right)+c_1^J+c_1^S+2\,\gamma_0^S\,\ln\frac{\mu}{p_T}+2\,\gamma_0^\text{cusp}\,C_S\,\ln^2\frac{\mu}{p_T},\\
p_1^{(1)}&=\gamma_0^J+2\gamma_0^S  +4\,\gamma_0^\text{cusp}\,C_S\,\ln\frac{\mu}{p_T},\\
p_2^{(1)}&=\gamma_0^\text{cusp}\left( C_J+4C_S \right).
\end{align}
The lengthy two-loop coefficients $p_i^{(2)}$ are listed in Appendix \ref{app:expansion}. The Casimir operators relevant for the different channels are
\begin{align} 
C_{S_{q\qb}}=C_F-\frac{C_A}{2}\,, \quad C_{S_{qg}}=\frac{C_A}{2} \,,\quad C_{S_{gg}}=\frac{C_A}{2}\,,\quad C_{J_g}=C_A\,, \quad C_{J_q}=C_F \,,
\end{align}
and the anomalous dimension coefficients are given by
\begin{align}
\gamma_0^{J_g}=-\beta_0\,, \quad\gamma_0^{J_q}=-3C_F\,,\quad \gamma_0^{S_{q\qb}}=0\,,\quad\gamma_0^{S_{qg}}=0\,,\quad\gamma_0^{S_{gg}}=0\,.
\end{align}
The nonlogarithmic one-loop coefficients of the gluon \cite{Becher:2009th} and quark \cite{Bauer:2003pi,Bosch:2004th} jet functions read
\begin{align}
c_1^{J_g}=C_A\left( \frac{67}{9}-\frac{2\pi^2}{3} \right)-\frac{20}{9}T_F \,n_f,\quad c_1^{J_q}=(7-\frac{2\pi^2}{3})C_F,
\end{align}
while the coefficients for the soft function read \cite{Becher:2009th}
\begin{align}
c_1^{S_{q\qb}}=\left( C_F-\frac{C_A}{2} \right)\pi^2\,, \quad c_1^{S_{qg}}=\frac{\pi^2 C_A}{2}\,, \quad c_1^{S_{gg}}=\frac{\pi^2 C_A}{2}\,.
\end{align}
The two-loop coefficient for the quark jet function has been calculated in \cite{Becher:2006qw}, for the gluon jet function in \cite{Becher:2010pd}, and  the two-loop coefficients for the soft functions have been calculated in \cite{Becher:2012za}; they are listed in Appendix \ref{app:jetsoft}.

\section{Numerical studies\label{sec:numerics}}

With all the ingredients in place, we now study the numerical size of the two-loop corrections. Before evaluating the full cross section, let us start by providing the numerical value of the two-loop hard functions at a fixed kinematic point. This only gives a rough estimate on the size of the corrections, but it also provides the reader with a numerical check should he or she implement the expressions obtained in the previous sections.
We choose $\hat{s}=1\, {\rm TeV}^2$, $\hat{t}=-0.4\, {\rm TeV}^2$ and $M_V=0.1\, {\rm TeV}$. These values imply that the transverse momentum is $p_T^2 = \hat{t}\hat{u}/\hat{s} \approx ( 0.5\, {\rm TeV})^2$. For the renormalization scale, we use $\mu = 0.6\, {\rm TeV}$ and obtain
\begin{align}
\hat{H}_{q\bar{q}\to V g}(\hat{u},\hat{t},\mu) &= 1+ \left( 1.47009 - 0.138371\,N_{V}^a \right) \alpha _s +   \left(3.89803\, -0.03923 \,N_{V}^v\right) \alpha _s^2\,,
\notag
\\
\hat{H}_{qg\to Vq }(\hat{u},\hat{t},\mu) &= 1 +\left(1.59193 + 0.114478 \,N_{V}^a \right) \alpha _s + \left(2.45463\, -0.02594\, N_{V}^v\right) \alpha _s^2
\,,\notag\\
H_{gg\to V g}(\hat{u},\hat{t},\mu) &= \left(g_s \,e\, Q_V^g \,d^{abc}\right)^2  1.19687 \, \alpha _s^2
\,.\label{eq:numhard}
\end{align}
To show the relative size of the corrections, we have normalized the hard functions for the $q\bar{q}$ and $qg$ channels to one at the lowest order, as indicated by the hat on the normalized functions. With $\alpha_s(\mu)\approx 0.09$, the corrections are moderate, of the order of a few per cent. The contribution proportional to $N_{V}^{v,a}$ arises from diagrams where the vector boson couples to an internal quark loop instead of the external quarks. By charge conservation, such contributions do not arise for $W$-bosons, so that $N^{v,a}_{W^\pm}=0$. For photons, we have
\begin{align}
N_{\gamma}^v &= \frac{1}{e_q}\sum_{q}e_q\,, & N_{\gamma}^a &= 0\,,
\end{align}
where the sum runs over the quark flavors in the loop and the denominator arises because we have factored out the charge in the definition of the hard function.
For $Z$-bosons, there are contributions from both the vector and the axial part of the coupling. The axial-vector part at one-loop order can be found in \cite{Gonsalves:1989ar}. In references \cite{Garland:2002ak,Gehrmann:2011ab} the two-loop vector part was computed, but the two-loop axial part is at present still unknown. The relevant couplings are
\begin{align}
N_{Z}^v & =  \frac{\left( L^Z_{qq}+R^Z_{qq} \right) }{|L^Z_{qq}|^2+|R^Z_{qq}|^2}\, \frac{1}{2} \sum_q (L_{qq}^V + R_{qq}^V)  \,, & N_{Z}^a & =  \frac{\left( L^Z_{qq}-R^Z_{qq} \right) }{|L^Z_{qq}|^2+|R^Z_{qq}|^2} \,\frac{1}{2} \sum_q (L_{qq}^V - R_{qq}^V)\,.
\end{align}
Note that the contributions proportional to $N^{v,a}_V$ are numerically very small. The normalization of the constant $N^{v}_Z$ differs from $N_{F,Z}$ in \cite{Garland:2002ak,Gehrmann:2011ab} because we consider the squared amplitude instead of the amplitude itself. 

Let us also evaluate the Higgs-boson hard functions with the same scale choice and the same kinematic point, with $M_H=0.1\, {\rm TeV}$. In this case, we find\footnote{In order to make the quark-channel amplitudes included in electronic form in the arXiv submission of \cite{Gehrmann:2011aa} consistent with the notation used in the paper, one has to change $p_1\leftrightarrow p_2$ and switch the sign of the amplitudes. We thank the authors for confirming this point.}
\begin{align}
\hat{H}_{gg\to H g}(\hat{u},\hat{t},\mu) &= 1+  6.02164 \,  \alpha _s +  24.2724  \,  \alpha _s^2\,, \notag\\
\hat{H}_{q \bar{q}\to H g}(\hat{u},\hat{t},\mu) &= 1+ 1.85023 \,\alpha _s +  8.15565 \, \alpha _s^2\,, \notag\\
 \hat{H}_{q g\to H q}(\hat{u},\hat{t},\mu) &=1 +  6.63865 \, \alpha _s + 24.9851  \,\alpha _s^2 \,.
\end{align}
These numbers do not include the small perturbative corrections to the Wilson coefficient $[C_t(m_t,\mu)]^2$ in \eqref{eq:Leff}. We observe that the higher-order terms are dramatically larger than in the vector-boson case, in line with the findings of \cite{Boughezal:2013uia}. Larger corrections are expected since the higher-order contributions to gluonic quantities are enhanced by factors of $C_A/C_F$ and also because the leading cross section is $\mathcal{O}(\alpha_s^3)$ for the Higgs case instead of $\mathcal{O}(\alpha_s)$ as in vector-boson production. However, for a meaningful assessment of the size of the higher-order corrections, one will need to evaluate the full cross section. Also, to make reliable predictions, one should check how large the corrections to the heavy top-quark limit are. The above values correspond to $p_T\approx 0.5\,{\rm TeV}$ for which the effective theory treatment is no longer appropriate. We will study the Higgs case in more detail in the future and will restrict ourselves to vector-boson production in the following.

\begin{table}[t!]
\centering
\begin{tabular}{lcccc cc}\toprule
\multirow{2}{*}{$\sigma(p_T>200\,{\rm GeV})\,[{\rm pb}]$} & \multicolumn{2}{c}{LHC at $7\,$TeV} & \multicolumn{2}{c}{LHC at $8\,$TeV}
& \multicolumn{2}{c}{LHC at $13\,$TeV}
\\ 
& $W^\pm$ &  $Z$ & $W^\pm$ &  $Z$ & $W^\pm$ &  $Z$
\\ \midrule
LO &
$34.6^{+6.3}_{-5.0}$ & $14.1^{+2.5}_{-2.0}$ & 
$47.4^{+8.1}_{-6.5}$ & $19.4^{+3.3}_{-2.7}$ &
$133^{+15}_{-18}$ & $55.7^{+7.3}_{-6.2}$
\\ 
NLO$_\mathrm{sing.}$
& $47.2^{+2.8}_{-3.1}$ & $19.2^{+1.1}_{-1.2}$ & 
$64.6^{+3.6}_{-4.0}$ & $26.5^{+1.4}_{-1.6}$ &
$181^{+8}_{-9}$ & $76.1^{+3.0}_{-3.5}$
\\ 
NNLO$_\mathrm{sing.}$
& $50.3^{+0.7}_{-0.5}$ & $20.5^{+0.2}_{-0.2}$ & 
$68.9^{+0.9}_{-0.6}$ & $28.3^{+0.3}_{-0.2}$ &
$194^{+2}_{-1}$ & $81.3^{+0.5}_{-0.1}$
\\ \midrule 
NLL &
$35.9^{+5.6}_{-4.8}$ & $14.7^{+2.3}_{-2.0}$ & 
$48.8^{+7.7}_{-6.6}$ & $20.1^{+3.2}_{-2.7}$ &
$133^{+22}_{-19}$ & $56.1^{+9.4}_{-8.2}$
\\ 
N$^2$LL &
$47.6^{+3.3}_{-2.9}$ & $19.4^{+1.4}_{-1.2}$ & 
$65.0^{+4.6}_{-4.1}$ & $26.7^{+1.9}_{-1.7}$ &
$180^{+15}_{-13}$ & $75.7^{+6.3}_{-5.4}$
\\ 
N$^3$LL &
$50.5^{+2.1}_{-1.1}$ & $20.6^{+0.9}_{-0.4}$ & 
$69.1^{+2.9}_{-1.5}$ & $28.3^{+1.3}_{-0.6}$ &
$193^{+10}_{-5}$ & $81.1^{+4.4}_{-2.3}$
\\ \midrule
NLO&
$53.5^{+5.2}_{-4.8}$ & $21.5^{+2.0}_{-1.9}$ & 
$73.5^{+7.0}_{-6.3}$ & $29.7^{+2.7}_{-2.5}$&
$209^{+19}_{-16}$ & $86.3^{+7.4}_{-6.4}$
\\ 
NNLO$_\mathrm{sing.}$+NLO &
$56.6^{+3.1}_{-2.2}$ & $22.8^{+1.1}_{-0.8}$ &
$77.8^{+4.3}_{-2.9}$ & $31.5^{+1.6}_{-1.1}$ &
$221^{+13}_{-7}$ & $91.5^{+4.9}_{-2.6}$ 
\\
N$^3$LL+NLO
& $56.8^{+2.2}_{-1.2}$ & $22.9^{+0.8}_{-0.4}$ & 
$77.9^{+3.1}_{-1.7}$ & $31.6^{+1.1}_{-0.6}$ &
$220^{+10}_{-6}$ & $91.3^{+3.7}_{-2.2}$ \\ \bottomrule
\end{tabular}
\vspace{-0.0cm}
\caption{The cross section $\sigma(p_T>200\,{\rm GeV})$ and its scale uncertainty using different approximations, see text. The columns labeled $W^\pm$ contain the result for the sum of the cross sections for $W^+$ and $W^-$ production.
\label{tab:values}}
\end{table}

In our previous work \cite{Becher:2011fc,Becher:2012xr}, we have used a {\sc Mathematica} code to compute the cross sections. With the large size of the expressions for the two-loop hard functions, this code becomes prohibitively slow and we have now developed a C\nolinebreak\hspace{-.05em}\raisebox{.4ex}{\tiny\textbf{+}}\nolinebreak\hspace{-.10em}\raisebox{.4ex}{\tiny\textbf{+}} code {\sc PeTeR} \cite{peter} to compute the cross section, which will be made public in the future. The code computes the resummed cross section near the partonic threshold as well as its fixed-order expansion. In addition, it also computes the full NLO fixed-order cross section.  
In a future paper, we will present a detailed phenomenological study of vector-boson production, including the two-loop corrections as well as electroweak Sudakov effects, which were recently treated using the same threshold-resummation framework \cite{Becher:2013zua}. For the moment, we        focus on the size of the two-loop QCD corrections and check how much they change the cross section. To do so, we  use the same input parameters as in our previous paper \cite{Becher:2012xr}, namely the NNLO MSTW 2008 PDF set and its associated $\alpha_s(M_Z) = 0.1171$~\cite{Martin:2009iq} with three-loop running, and $M_Z = 91.1876$ GeV, $M_W = 80.399$ GeV, $\alpha_{\rm e.m.} = 127.916^{-1}$, $\sin^2\theta_W = 0.2226$, $|V_{ud}| = 0.97425$, $|V_{us}| = 0.22543$, $|V_{ub}| = 0.00354$, $|V_{cd}| = 0.22529$, $|V_{cs}| = 0.97342$, $|V_{cb}| = 0.04128$.  We treat all partons as massless except for the top quark, which is integrated out from the theory. The hard function for $Z$-boson production receives tiny contributions from the axial-vector coupling, see \eqref{eq:numhard}. At one-loop order they are due to triangle diagrams. A similar contribution is present for the $gg$ channel \cite{vanderBij:1988ac,Hopker:1993pb}, which is not included so far but might be of a similar order of magnitude as the NLO triangle contribution. For simplicity, and because the two-loop axial corrections are not known, we set $N^{v}_{V}=0$. Numerically the two-loop $N^{v}_{V}$ terms are negligibly small. 

\begin{figure}[t!]
\centering
\includegraphics[width=0.45\textwidth]{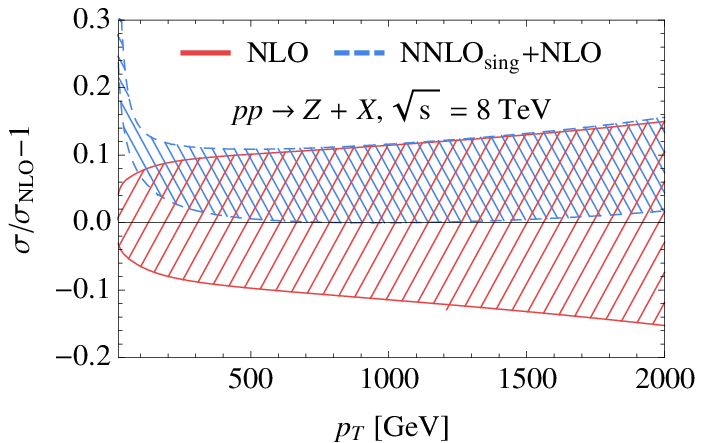}
\hspace{0.02\textwidth}
\includegraphics[width=0.45\textwidth]{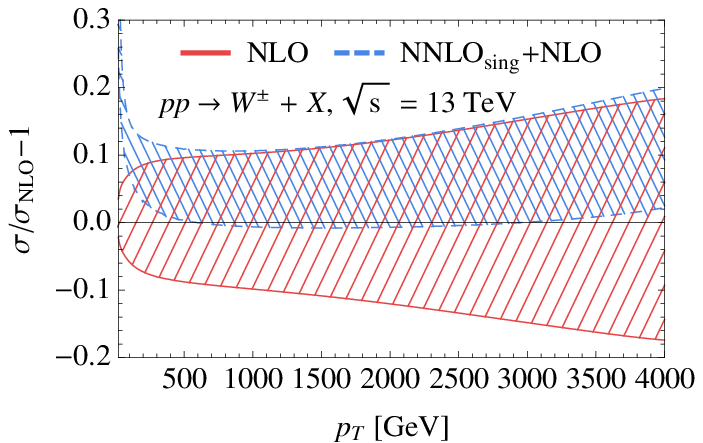}
\caption{Scale uncertainty bands relative to the NLO result for $Z$ production at the LHC with $\sqrt{s}=8\,{\rm TeV}$ (left) and for combined $W^\pm$ production at the LHC with $\sqrt{s}=13\,{\rm TeV}$ (right) for NLO and NNLO$_\text{sing}+$NLO.}
\label{fig:ErrorBands}
\end{figure}

A list of values for the integrated cross section $\sigma(p_T>200\,\text{GeV})$ is shown in Table~\ref{tab:values} for different LHC center-of-mass energies. The table presents three different approximations i) the fixed-order threshold cross section ii) the resummed results, and iii) the results obtained after matching to the known NLO fixed-order result. The entries LO, NLO$_{\rm sing.}$, NNLO$_{\rm sing.}$ show the perturbative expansion of the threshold cross section, which consists of the singular distributions defined in  (\ref{eq:star}). Since the LO partonic cross section is proportional to $\delta(m_X^2)$ it is purely singular. Beyond leading order, the cross section also has regular pieces not associated with soft and collinear radiation. As the table shows, the regular pieces obtained from the difference NLO$-$NLO$_{\rm sing}$ are of moderate size. For example, for $Z$-production at $\sqrt{s}=8\,{\rm TeV}$, the singular pieces amount to about 70\% of the NLO correction. The fact that the singular pieces amount to the bulk of the cross section is true in many other cases as well, and we therefore expect that the singular pieces will provide a good approximation to the full NNLO correction. The column NNLO$_{\rm sing.}$+NLO shows the result obtained if both the full NLO result and the singular pieces at NNLO are included. For the factorization and renormalization scales, we use
 \begin{equation}\label{muh}
\mu= \mu_r=\mu_f=\frac{13p_T+2M_V}{12}-\frac{p_T^2}{\sqrt{s}}\,.
 \end{equation}
 This value is close to $p_T$ and was adopted as the default scale  $\mu_h$ for the hard function after a numerical study in \cite{Becher:2012xr}. The scale uncertainty is obtained by varying the scale $\mu$ by  a factor two around the default value. Figure~\ref{fig:ErrorBands} shows the resulting uncertainty bands for $Z$ and for $W^+$ or $W^-$ production at NLO and NNLO$_\text{sing}$+NLO. The results are normalized to the NLO result at the default scale choice. We find that including the two-loop singular terms corresponds to a shift of about $+5\%$ of the cross section and decreases the scale uncertainty by a factor of two, compared to NLO.
 
 \begin{figure}[t!]
\centering
\includegraphics[width=0.46\textwidth]{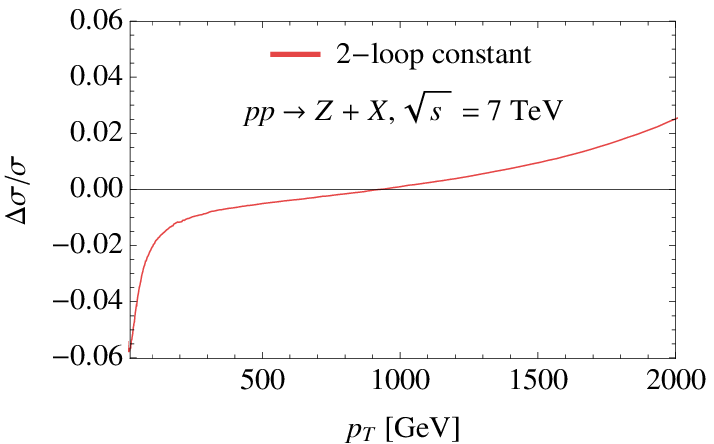}
\hspace{0.02\textwidth}
\includegraphics[width=0.45\textwidth]{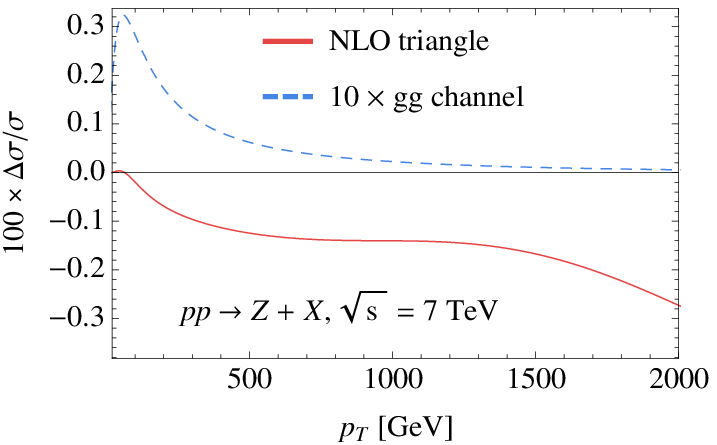}
\caption{Left: Relative change of the NNLO$_\text{sing}$ cross section compared with an approximation used in \cite{Becher:2011fc,Becher:2012xr}, in which the two-loop constant was chosen to vanish at $\mu=p_T$.
Right: Relative contribution of triangle diagrams at NLO (red line) and of the gluon-gluon channel at NNLO (times 5, blue dashed line).}
\label{fig:GGTriangle}
\end{figure}

The factorization formula (\ref{eq:factorization}) can be used to resum the singular pieces to all orders using RG evolution in SCET. For $W$ and $Z$ production, this was done in \cite{Becher:2011fc,Becher:2012xr}. To perform the resummation, we evolve the hard, jet and soft functions from their characteristic scales to the factorization scale.  We adopt here the same default scales as in the papers \cite{Becher:2011fc,Becher:2012xr}. Resummation to N$^{n+1}$LL accuracy requires the hard, jet and soft functions at N$^n$LO. Comparing the resummed N$^{n+1}$LL and the threshold fixed-order results N$^n$LO$_{\rm sing.}$, we find that they are numerically very similar. Resummation is thus not a large effect, since the characteristic scales for the jet and soft functions are not much below the hard scale. Their numerical values depend on the fall-off of the PDFs towards larger $x$, which enhances the threshold region. After a numerical study, following \cite{Becher:2007ty}, the values 
\begin{equation}\label{muj}
\mu_j= \frac{p_T}{2}\left(1-\frac{p_T}{\sqrt{s}}\right)\,
 \end{equation}
 and $\mu_s = \mu_j^2/\mu_h$ were adopted in \cite{Becher:2012xr}. Since the numerical values of the jet and soft scales are not much lower than the hard scale, the logarithms which are resummed are of moderate size. 

 The highest order result obtained in \cite{Becher:2011fc,Becher:2012xr} was denoted by N$^3$LL$_{\rm p}$ where the label ``p'' (for partial) indicated that the two-loop hard function was missing. With this ingredient in place, our results now have full N$^3$LL accuracy.\footnote{Strictly speaking, there is one more unknown ingredient, namely the four-loop cusp anomalous dimension, but its numerical impact is negligibly small.} However, since all the logarithmic pieces of the hard function follow from RG-invariance, they were already included in N$^3$LL$_{\rm p}$ of \cite{Becher:2011fc,Becher:2012xr}. The logarithms were introduced such that their contribution vanishes at $\mu=p_T$, i.e.\ the two-loop constant was defined as the value of the hard function at $\mu = p_T$. To see how much the results in \cite{Becher:2011fc,Becher:2012xr} change due to presence of the two-loop constant, we plot its value compared to the cross section without the two-loop constant. The size of the effect depends on the transverse momentum. As shown in the left panel of Figure \ref{fig:GGTriangle}, it reduces the cross section by about two per cent at $p_T \approx 100\, {\rm GeV}$ and enhances it by a similar amount at large $p_T \approx 2\, {\rm TeV}$. The change is within the scale uncertainties of the N$^3$LL$_{\rm p}$ results and smaller than the total two-loop effect, which is of order $+5\%$, see Figure \ref{fig:ErrorBands}.

 There are two more effects, which we briefly address. The first is the contribution of the gluon-gluon channel in $Z$ production, which first arises at NNLO and is shown by the blue dashed line in the right panel of Figure~\ref{fig:GGTriangle}. The channel gives a very small positive contribution to the cross section. It peaks around $p_T=50$\,GeV and is smaller than 1\textperthousand. The second effect, shown by the red line in the right panel of the figure is the triangle contribution at NLO which arises due to the axial coupling of the $Z$ boson. This contribution is mostly negative and smaller than 3\textperthousand. As we stressed earlier, there are also axial contributions at NNLO, in particular also in the gluon-gluon channel, which were obtained in \cite{vanderBij:1988ac,Hopker:1993pb} but are not included here. According to these papers, the axial corrections to the gluon-gluon channel are bigger than the vector contributions and could be comparable in size to the (very small) axial contributions at NLO.
 
\section{Conclusion}

Radiative corrections to hard-scattering processes simplify considerably near the partonic threshold. In this work, we have used these simplifications to obtain the NNLO corrections to transverse-momentum spectra of photons, $W$, $Z$ and Higgs bosons. Our results are valid at large transverse momentum $p_T$ of the electroweak boson, where the invariant mass of the recoiling jet is small compared to $p_T$. As the threshold terms often capture the bulk of the radiative corrections, we expect that our results are a good approximation to the exact NNLO results. In addition, the threshold terms can serve as a check on the full NNLO results, once they become available.

The starting point of our analysis is the threshold factorization of the cross section into hard, jet and soft functions. Building on earlier work, in which we computed the two-loop collinear and soft functions, we computed the last missing NNLO ingredient, the hard functions, in this work. To this end, we converted known results for on-shell $V$+jet  amplitudes into $\overline{\text{MS}}$-subtracted hard functions defined in Soft-Collinear Effective Theory. The conversion procedure presented in Sections \ref{sec:hard} and \ref{sec:helicity} is completely general, and applies similarly to other processes. Our calculation also provides the last missing ingredient to resum the threshold terms to N$^3$LL accuracy. 

We have implemented the NNLO threshold corrections and the N$^3$LL resummed results into a C\nolinebreak\hspace{-.05em}\raisebox{.4ex}{\tiny\textbf{+}}\nolinebreak\hspace{-.10em}\raisebox{.4ex}{\tiny\textbf{+}} code {\sc PeTeR} \cite{peter}, which will be made public in the future. For $W$ and $Z$ production, we find that the NNLO threshold corrections are moderate. They enhance the cross section by about $5\%$, and they reduce the scale uncertainty by about a factor of two. In addition, we have also given resummed results at N$^3$LL accuracy, matched to NLO fixed-order results. Numerically, we find that the resummation effects, i.e. terms beyond NNLO, are not very important. Our final results for the integrated cross sections with  $p_T >200$~GeV are given in the last two lines of Table \ref{tab:values}. For Higgs production, the corrections to the hard functions are much larger than in the vector-boson case and resummation will likely be more important. We will present numerical results for the Higgs cross section in the future.

 For an accurate description of LHC data for vector-boson production at high-$p_T$, one also needs to implement electroweak corrections which are large and negative. These Sudakov-type corrections have recently been studied using the same threshold-resummation formalism. In a next step, we will perform a detailed phenomenological analysis of vector-boson production, including both electroweak and QCD corrections.  The hard functions determined in the present paper are relevant not only for hadronically inclusive boson production, but are also needed for resummations of more exclusive one-jet observables such as jet-mass spectra or jet-veto cross sections.
 
\vspace{0.6cm}
{\em Acknowledgments:\/}
We thank T.\ Gehrmann for discussions and help with the two-loop helicity amplitudes.
T.B.\ is grateful to KITP Santa Barbara for hospitality and support. T.B.\ and G.B.\ would like to thank the ESI Vienna for hospitality and support. We thank Fabrizio Caola, Frank Petriello and Markus Schulze for pointing out a numerical problem in the quark-gluon channel in the published version of this paper and Thomas Gehrmann and Matthieu Jacquier for providing a corrected helicity amplitude for this channel. The work of T.B.\ is supported by the Swiss National Science Foundation (SNF) under grant 200020-140978. G.B.\ gratefully acknowledges the support of a University Research Fellowship by the Royal Society.

\begin{appendix}
\numberwithin{equation}{section}

\section{Anomalous dimensions and IR-subtraction terms}\label{renapp}

In the following, we give explicit expressions for the two-loop coefficients needed for the renormalization of the hard function. For all the anomalous dimensions below also the three-loop result is known and can be found, for example in \cite{Becher:2009qa}. 

We define the QCD $\beta$-function and its expansion as
\begin{align}
\beta(\alpha_s)&=
\frac{d\alpha_s}{d\ln\mu} 
=-2\alpha_s\left[  \left( \frac{\alpha_s}{4\pi}\right)\beta_0+\left( \frac{\alpha_s}{4\pi}\right)^2\beta_1+\dots \right] \,,
\end{align}
so that the lowest two coefficients have the explicit form
\begin{align}
\beta_0&=\frac{11}{3}C_A-\frac{4}{3}\,T_F\,n_f\, , &
\beta_1&=\frac{34}{3}\,C_A^2-\frac{20}{3}\,C_A\,T_F\,n_f-4\,C_F\,T_F\,n_f\,. \nonumber
\end{align}
In the following, we will expand all anomalous dimensions in units of $\alpha_s/4\pi$, and we denote the expansion coefficients in the form
\begin{align}
\gamma(\alpha_s) &= \left( \frac{\alpha_s}{4\pi}\right) \gamma_0 +\left( \frac{\alpha_s}{4\pi}\right)^2 \gamma_1+ \dots \;.
\end{align}
Up to two-loop order, the cusp anomalous dimension $\gamma^\text{cusp}$ is given by 
\begin{align}
\gamma_0^\text{cusp}&=4\,, &
\gamma_1^\text{cusp}&=\left(\frac{268}{9}-\frac{4\pi^2}{3}   \right) C_A-\frac{80}{9}\,T_F\,n_f \,,
\end{align}
and the collinear anomalous dimensions $\gamma^q$ and $\gamma^g$ are
\begin{align}
  \gamma_0^q &= -3 C_F \,, \nonumber\\
   \gamma_1^q &= C_F^2 \left( -\frac{3}{2} + 2\pi^2
    - 24\zeta_3 \right)
    + C_F C_A \left( - \frac{961}{54} - \frac{11\pi^2}{6} 
    + 26\zeta_3 \right)
    + C_F T_F n_f \left( \frac{130}{27} + \frac{2\pi^2}{3} \right) \, ,  \nonumber \\
   \gamma_0^g &= - \beta_0  \,, \\
   \gamma_1^g &= C_A^2 \left( -\frac{692}{27} + \frac{11\pi^2}{18}
    + 2\zeta_3 \right) 
    + C_A T_F n_f \left( \frac{256}{27} - \frac{2\pi^2}{9} \right)
    + 4 C_F T_F n_f  \,.    \nonumber
\end{align}

The renormalization factor $\boldsymbol{Z}$ is obtained by solving its RG equation, which is driven by the anomalous dimension matrix $\boldsymbol{\rG}$ in (\ref{eq:hardanomdim}). 
The two-loop expression has the form
\begin{align}\label{Zfact}
\ln\boldsymbol{Z}=\frac{\alpha_s}{4\pi}\left[ \frac{\rG'_0}{4\epsilon^2} + \frac{\boldsymbol{\rG}_0}{2\epsilon} \right]+\left(\frac{\alpha_s}{4\pi}\right)^2\left[ -\frac{3\beta_0\rG'_0}{16\epsilon^3}+\frac{\rG'_1-4\beta_0\boldsymbol{\rG}_0}{16\epsilon^2} +\frac{\boldsymbol{\rG_1}}{4\epsilon}\right]+\mathcal{O}(\alpha_s^3)
\end{align}
with
\begin{align}
\rG'(\alpha_s)\equiv\frac{\partial}{\partial\ln\mu}\boldsymbol{\rG}(\{p\},\mu)=-\gamma_{\text{cusp}}(\alpha_s)\sum_i C_i\,.
\end{align}
Expanding the {\em inverse} $\boldsymbol{Z}$-factor in units of $\alpha_s/2\pi$,
\begin{align}\label{eq:Zfact}
\boldsymbol{Z}^{-1}(\epsilon,\,\{p\},\,\mu)&=
1+\frac{\alpha_s}{2\pi} \,\bm{Z}^{(1)}(\eps)
+\left(\frac{\alpha_s}{2\pi}\right)^2 \bm{Z}^{(2)}(\eps)
+\mathcal{O}(\alpha_s^3) \,,
\end{align}
one obtains 
\begin{align}
\bm{Z}^{(1)}(\eps) &= -\frac{\rG'_0}{8\epsilon^2} - \frac{\boldsymbol{\rG}_0}{4\epsilon}  \,,\nonumber \\
\bm{Z}^{(2)}(\eps) &=  \frac{\rG'^2_0}{128\epsilon^4}+\frac{3\beta_0\rG'_0+2\rG'_0\boldsymbol{\rG}_0}{64\epsilon^3}
+\frac{4\beta_0\boldsymbol{\rG}_0+2\boldsymbol{\rG}_0^2-\rG'_1}{64\epsilon^2}-\frac{\boldsymbol{\rG}_1}{16\epsilon}\, .
\end{align}
The one-loop subtraction operator appearing in Catani's formula for the IR divergences is 
\begin{align} \label{eq:catani1}
\boldsymbol{I}^{(1)}(\epsilon)&=\frac{e^{\epsilon \gamma_E}}{\rG(1-\epsilon)}\sum_i \left( \frac{1}{\epsilon^2}-\frac{\gamma_0^i}{2\epsilon}\frac{1}{C_i} \right)\sum_{j\neq i}\frac{\boldsymbol{T}_i\cdot\boldsymbol{T}_j}{2}\left( \frac{\mu^2}{-s_{ij}} \right)^\epsilon 
\\
&\equiv 
\frac{\rG'_0}{8\epsilon^2} + \frac{\boldsymbol{\rG}_0}{4\epsilon} 
+\sum_{n=0}^\infty \boldsymbol{\mathcal{C}_n} \eps^n
\,.
\end{align}
Apart from the pole terms, we will need the explicit expressions for the first two coefficients
\begin{align}
\boldsymbol{\mathcal{C}_0}
&=\sum_{(i,j)}\frac{\bT_i\cdot \bT_j}{16} 
\left[\gamma_0^{\text{cusp}} \, \ln^2\frac{\mu^2}{-s_{ij}}
- \frac{4\gamma_0^i}{C_i} \, \ln\frac{\mu^2}{-s_{ij}} \right]
-\frac{\pi^2}{96} \rG'_0
\,,
\end{align}
and 
\begin{align}
\boldsymbol{\mathcal{C}_1}
&=\sum_{(i,j)}\frac{\bT_i\cdot \bT_j}{48} 
\left[\gamma_0^{\text{cusp}} \, \ln^3\frac{\mu^2}{-s_{ij}}
- \frac{6\gamma_0^i}{C_i} \, \ln^2\frac{\mu^2}{-s_{ij}} \right]
-\frac{\pi^2}{48} \boldsymbol{\rG}_0 
-\frac{\zeta_3}{24} \rG'_0
\,.
\end{align}
The two-loop subtraction operator is defined as
\begin{align} \label{eq:catani2}
\boldsymbol{I}^{(2)}(\epsilon)&=\frac{e^{-\epsilon \gamma_E}\rG(1-2\epsilon)}{\rG(1-\epsilon)}\left( \frac{\gamma_1^{\text{cusp}}}{8}+\frac{\beta_0}{2\epsilon} \right)\bI^{(1)}(2\epsilon)-\frac{1}{2}\bI^{(1)}(\epsilon)\left( \bI^{(1)}(\epsilon)+\frac{\beta_0}{\epsilon} \right)+\boldsymbol{H}^{(2)}_{\text{R.S.}}(\epsilon) \,,
\end{align}
where the last term has not been specified in \cite{Catani:1998bh}, but is was stated that it only contains single poles. Using the expression for the $\bm{Z}$-factor \eqref{eq:Zfact} one can derive this term. Explicitly, we find
\begin{align}\label{eq:H2RS}
\boldsymbol{H}^{(2)}_{\text{R.S.}}(\epsilon) 
&=
\frac{if^{abc}}{384\epsilon} \,(\gamma_0^\text{cusp})^2
\sum_{(i,\,j,\,k)}\bT_i^a\,\bT_j^b\,\bT_k^c\;
 \ln\frac{-s_{ij}}{-s_{jk}}\,\ln\frac{-s_{jk}}{-s_{ki}}\,\ln\frac{-s_{ki}}{-s_{ij}}
\nonumber\\
&\phantom{=\;}
- \frac{if^{abc}}{128\epsilon} \,\gamma_0^\text{cusp}
\sum_{(i,\,j,\,k)}\bT_i^a\,\bT_j^b\,\bT_k^c\;
\left( \frac{\gamma_0^i}{C_i} - \frac{\gamma_0^j}{C_j} \right)
\ln\frac{-s_{ij}}{-s_{jk}}\,\ln\frac{-s_{ki}}{-s_{ij}}
\nonumber\\
&\phantom{=\;}
+ \frac{\boldsymbol{\rG}_1}{16\epsilon}
- \frac{\gamma_1^\text{cusp} \, \boldsymbol{\rG}_0}{64\epsilon} 
-\frac{\pi^2 \beta_0 \rG'_0}{256\epsilon} 
\,,
\end{align}
where the two sums run over all unordered triplets of distinct parton indices. The terms in the first two lines are equal to $\frac{1}{8}\big[ \bm{\Gamma_0},\bm{\mathcal{C}_0} \big]$. This commutator can be simplified by noting that the contributions which involve four different partons vanish because the color generators associated with different partons commute. This expression for $\boldsymbol{H}^{(2)}_{\text{R.S.}}(\epsilon) $ was derived in \cite{Becher:2009cu,Becher:2009qa}, but the term in the second line, which involves the collinear anomalous dimensions $\gamma_0^i$ was missed. This extra contribution was discussed in Appendix D of \cite{Aybat:2006mz}, where it was shown that it can only contribute for amplitudes with more than four external particles. 

The two-loop conversion relation in (\ref{eq:twoloopconv}) involves a commutator of the one-loop anomalous dimension $\bm{\Gamma_0}$ with the $\mathcal{O}(\epsilon)$ term in the expansion of $\boldsymbol{I}^{(1)}(\epsilon)$. This commutator can be simplified to 
\begin{align}
&\big[ \bm{\Gamma_0} ,\bm{\mathcal{C}_1}  \big]  
=
\frac{if^{abc}}{144} \,(\gamma_0^\text{cusp})^2
\sum_{(i,\,j,\,k)}\bT_i^a\,\bT_j^b\,\bT_k^c\;
 \ln\frac{-s_{ij}}{-s_{jk}}\,\ln\frac{-s_{jk}}{-s_{ki}}\,\ln\frac{-s_{ki}}{-s_{ij}} \,
 \ln \frac{\mu^6}{(-s_{ij})(-s_{jk})(-s_{ki})}
\nonumber\\
&\hspace{1.9cm}
- \frac{if^{abc}}{48} \,\gamma_0^\text{cusp}
\sum_{(i,\,j,\,k)}\bT_i^a\,\bT_j^b\,\bT_k^c\;
\left\{
\left( \frac{\gamma_0^i}{C_i} - \frac{\gamma_0^j}{C_j} \right)
\ln\frac{-s_{ij}}{-s_{jk}}\,\ln\frac{-s_{ki}}{-s_{ij}} \,\ln \frac{\mu^6}{(-s_{ij})(-s_{jk})(-s_{ki})}
 \right.
 \nonumber\\
&\hspace{6.9cm}\left.
+ \left( \frac{\gamma_0^i}{C_i} + \frac{\gamma_0^j}{C_j} \right)
 \ln\frac{-s_{ij}}{-s_{jk}}\,\ln\frac{-s_{jk}}{-s_{ki}}\,\ln\frac{-s_{ki}}{-s_{ij}}
 \right\}\,.
\end{align}

\section{Hard, jet and soft functions at two-loop order\label{app:jetsoft}}

\subsection{Hard function}

The helicity amplitudes in \cite{Garland:2002ak,Gehrmann:2011ab,Gehrmann:2013vga,Gehrmann:2011aa} were only given for the choice $\mu^2 = q^2$. The full $\mu$ dependence can be reconstructed by solving the associated RG equation \cite{Becher:2009cu}, driven by the anomalous dimension $\boldsymbol{\rG}$ in \eqref{eq:hardanomdim}. One finds \cite{Becher:2009th}
\begin{align}
\hat{H} \left( \hat{u}, \hat{t},\mu \right) &= 1
+ \left( \frac{\alpha_s}{4 \pi} \right) \left\{ -\Gamma^H_0 \frac{L^2}{2} - \gamma_0^H L + c_1^H \right\}
\\
& + \left( \frac{\alpha_s}{4 \pi} \right)^2 
\bigg\{ \left( \Gamma_0^H  \right)^2 \frac{L^4}{8}
+  \left( \beta_0+   3 \gamma_0^H \right) \Gamma_0^H \frac{L^3}{6}
\nonumber\\
& \qquad + \left[ \gamma_0^H(\beta_0+\gamma_0^H) - \Gamma_1^H - \Gamma_0^H c_1^H \right] \frac{L^2}{2}
\nonumber \\
& \qquad +
\left[ - c_1^H (\beta_0 + \gamma_0^H) - \gamma_1^H \right] L + c_2^H \bigg\} 
+\mathcal{O}(\alpha_s^3)\nonumber\,.
\end{align}
The logarithms for the different channels are
\begin{align}
L_{q\bar{q}} &= L_{gg} = \ln\frac{\hat{s}}{\mu^2} \;, \quad & L_{qg} &= \ln\frac{-\hat{u}}{\mu^2} \,.
\end{align}
The anomalous dimensions can be extracted from the general result \cite{Becher:2009qa}, explicitly,
\begin{align}
\Gamma^{H_{q\bar{q}}}(\alpha_s) &= \Gamma^{H_{qg}}(\alpha_s) = \left( C_F + \frac{C_A}{2} \right) \gamma_{\mathrm{cusp}}(\alpha_s) \,, \nonumber\\
\Gamma^{H_{gg}}(\alpha_s) &= \frac{3C_A}{2} \gamma_{\mathrm{cusp}}(\alpha_s) \,,
\nonumber \\
\gamma^{H_{q\bar{q}}}(\alpha_s) &= 2\gamma^q(\alpha_s)+\gamma^g(\alpha_s)-\frac{C_A}{2} \gamma_{\mathrm{cusp}}(\alpha_s) \ln\frac{\hat{s}^2}{\hat{t}\hat{u}} - \frac{\beta(\alpha_s)}{2\alpha_s} \,,
\nonumber \\
\gamma^{H_{qg}}(\alpha_s) &=  2\gamma^q(\alpha_s)+\gamma^g(\alpha_s)-\frac{C_A}{2} \gamma_{\mathrm{cusp}}(\alpha_s) \ln\frac{\hat{u}^2}{-\hat{s}\hat{t}} - \frac{\beta(\alpha_s)}{2\alpha_s} \,,
\nonumber \\
\gamma^{H_{gg}}(\alpha_s) &= 3\gamma^g(\alpha_s)-\frac{C_A}{2} \gamma_{\mathrm{cusp}}(\alpha_s) \ln\frac{\hat{s}^2}{\hat{t}\hat{u}} - \frac{3\beta(\alpha_s)}{2\alpha_s} \,.
\end{align}
The results for the other crossed channels like $gq$ can be obtained as usual by replacing $\hat{t}\leftrightarrow \hat{u}$. 

For Higgs production, the last term of $\gamma^{H_{q\bar{q}}}(\alpha_s) $ and $\gamma^{H_{qg }}(\alpha_s)$ must be changed to  $- \frac{3\beta(\alpha_s)}{2\alpha_s}$ because the leading-order cross sections start at $\mathcal{O}(\alpha_s^3)$. Also, the above result is relevant for the full hard function, which includes the factor $[C_t(m_t^2,\mu^2)]^2$, from the Wilson coefficient of the operator \eqref{eq:Leff} which mediates Higgs production in the large-$m_t$ limit. The NNLO value of this coefficient is \cite{Kramer:1996iq,Chetyrkin:1997iv}
\begin{eqnarray}\label{Ct}
   C_t(m_t^2,\mu^2)
   &=& 1 + \frac{\alpha_s}{4\pi}\,(5C_A-3C_F) \nonumber\\
   &&\mbox{}+ \left( \frac{\alpha_s}{4\pi} \right)^2
    \bigg[ \frac{27}{2}\,C_F^2 
    + \left( 11\ln\frac{m_t^2}{\mu^2} - \frac{100}{3} \right) C_F C_A 
    - \left( 7\ln\frac{m_t^2}{\mu^2} - \frac{1063}{36} \right) C_A^2 
    \nonumber\\
   &&\quad\mbox{}- \frac{4}{3}\,C_F T_F - \frac{5}{6}\,C_A T_F 
   - \left( 8\ln\frac{m_t^2}{\mu^2} + 5 \right) C_F T_F n_f 
   - \frac{47}{9}\,C_A T_F n_f \bigg] \,.
\end{eqnarray} 
 If the scale-dependent factor $[C_t(m_t^2,\mu^2)]^2$ is divided out, the anomalous dimension of the hard function changes by $\Delta\gamma^{H_{ab}}(\alpha_s) = -2 \gamma_t(\alpha_s)$, where 
\begin{equation}\label{eq:rgech}
   \frac{d}{d\ln\mu}\,C_t(m_t^2,\mu^2) 
   = \gamma^t(\alpha_s)\,C_t(m_t^2,\mu^2) \,,
    \qquad \mbox{with} \quad
   \gamma^t(\alpha_s) 
   = \alpha_s^2\,\frac{d}{d\alpha_s}\,
   \frac{\beta(\alpha_s)}{\alpha_s^2} \,.
\end{equation}
The anomalous dimension $\gamma_t(\alpha_s)$ is related to the QCD $\beta$-function \cite{Inami:1982xt,Grinstein:1988wz} since the operator is proportional to the Yang-Mills Lagrangian.

\subsection{Jet function}
The expression for the Laplace-transformed jet function $\tilde{j}_c(L,\,\mu)$, with $c=q$ or $c=g$, reads
\begin{align}\label{eq:laplaceJet}
\tilde{j}_c(L,\mu)=1&+\frac{\alpha_s}{4\pi}\bigg[\frac{\Gamma_0^{J_c}}{2}L^2+\gamma_0^{J_c}\,L+ c_1^{J_c}  \bigg]
\nonumber
\\
&+\left(\frac{\alpha_s}{4\pi} \right)^2\bigg[\frac{(\Gamma_0^{J_c})^2}{8}\,L^4 +\frac{\Gamma_0^{J_c}}{6}\left(3 \gamma_0^{J_c}-\beta_0 \right)L^3  
+\frac{1}{2}\left(\gamma_0^{J_c}(\gamma_0^{J_c}-\beta_0)+c_1^{J_c}\Gamma_0^{J_c}+\Gamma_1^{J_c}  \right)L^2   \nonumber
\\
& \phantom{\left(\frac{\alpha_s(\mu)}{4\pi} \right)^2}\quad\;\;\,+\left( c_1^{J_c}(\gamma_0^{J_c}-\beta_0)+\gamma_1^{J_c} \right)L  +c_2^{J_c}  \bigg]   + \mathcal{O}(\alpha_s^3) \,.
\end{align}
This expression is obtained by solving the associated RG equation, which is governed by the anomalous dimensions
\begin{align}
   \Gamma^{J_q}(\alpha_s)&=C_F\,\gamma_\text{cusp}(\alpha_s)\,, \nonumber\\
   \gamma_0^{J_q} &= - 3 C_F  \, ,\\
\gamma_1^{J_q} &= C_F^2 \left( - \frac{3}{2} + 2 \pi^2 - 24 \zeta_3 \right) +
C_F C_A \left( - \frac{1769}{54} - \frac{11 \pi^2}{9} + 40 \zeta_3 \right) +
C_F T_F n_f \left( \frac{242}{27} + \frac{4 \pi^2}{9} \right) \, , \nonumber
\end{align}
in the quark case and
\begin{align}
   \Gamma^{J_g}(\alpha_s)&=C_A\,\gamma_{\text{cusp}}(\alpha_s)\,, \nonumber \\ 
   \gamma_0^{J_g} &= - \beta_0\,,  \\ 
   \gamma_1^{J_g} &= C_A^2 \left( - \frac{1096}{27} + \frac{11 \pi^2}{9} + 16  \zeta_3 \right) + C_A n_f T_F \left( \frac{368}{27} - \frac{4 \pi^2}{9}  \right) + 4 C_F T_F n_f  ,\nonumber 
\end{align}
for the gluon jet function. The nonlogarithmic coefficients are
\begin{align}
c_1^{J_q}&=C_F\left( 7-\frac{2\pi^2}{3}  \right) \,,
\nonumber \\
c_1^{J_g}&=C_A\left( \frac{67}{9}-\frac{2\pi^2}{3}  \right)-\frac{20}{9}\, T_F\,n_f \,,
\nonumber \\
c_2^{J_q}&=C_F^2 \left( \frac{205}{8} - \frac{97 \pi^2}{12} + \frac{61
  \pi^4}{90} - 6 \zeta_3 \right) + C_F C_A \left( \frac{53129}{648} -
  \frac{155 \pi^2}{36} - \frac{37 \pi^4}{180} - 18 \zeta_3 \right) \nonumber\\
 &\hspace{1cm} + C_F T_F n_f  \left( - \frac{4057}{162} + \frac{13 \pi^2}{9} \right)\,,\\
c_2^{J_g}&=C_A^2\left( \frac{20215}{162} -\frac{362\pi^2}{27}-\frac{88\,\zeta_3}{3}+\frac{17\pi^4}{36}  \right)+C_A\,T_F\,n_f\left( -\frac{1520}{27}+\frac{134\pi^2}{27}-\frac{16\,\zeta_3}{3}  \right)\nonumber\\
&\hspace{1cm} +C_F\,T_F\,n_f\left( -\frac{55}{3}+16\,\zeta_3 \right)+T_F^2\,n_f^2\left( \frac{400}{81}-\frac{8\pi^2}{27}   \right) \,. \nonumber
\end{align}

\subsection{Soft function}
The Laplace-transformed soft function reads
\begin{multline}\label{eq:laplaceSoft}
\tilde{s}_{ab}(L,\,\mu)=1+\frac{\alpha_s}{4\pi}\bigg[ 2\Gamma_0^{S_{ab}}\,L^2+2\gamma_0^{S_{ab}} L+c_1^{S_{ab}}   \bigg] \\
+\left( \frac{\alpha_s}{4\pi}  \right)^2\bigg[ 2(\Gamma_0^{S_{ab}})^2L^4+\frac{4\Gamma_0^{S_{ab}}}{3}(3\gamma_0^{S_{ab}}-\beta_0)\,L^3  +2\left( \gamma_0^{S_{ab}}\left(\gamma_0^{S_{ab}}-\beta_0\right)+\Gamma_0^{S_{ab}} c_1^{S_{ab}} +\Gamma_1^{S_{ab}} \right)L^2  \\
+2\left( c_1^{S_{ab}} \left(\gamma_0^{S_{ab}}-\beta_0\right)+ \gamma_1^{S_{ab}} \right)L+c_2^{S_{ab}}\bigg]
+\mathcal{O}(\alpha_s^3)\,.
\end{multline}
The anomalous dimensions in the above expression are
\begin{align}
\Gamma^{S_{ab}}(\alpha_s)&=C_{S_{ab}}\,\gamma_{\text{cusp}}(\alpha_s)\,,\nonumber \\
\gamma_0^{S_{ab}}&=0\,,\\
\gamma_1^{S_{ab}}&=C_{S_{ab}}\left( \left( 28\,\zeta_3-\frac{808}{27}+\frac{11\pi^2}{9} \right)C_A  +\left( \frac{224}{27}-\frac{4\pi^2}{9}  \right) T_F\,n_f  \right)\, , \nonumber
\end{align}
and the constants are given by
\begin{align}
c_1^{S_{ab}}=\,&C_{S_{ab}}\,\pi^2 \,,
\nonumber \\
c_2^{S_{ab}}=\,&\frac{1}{2}\left(C_{S_{ab}} \pi^2\right)^2+C_{S_{ab}}\,C_A\left(  \frac{2428}{81}+\frac{335\pi^2}{54}-\frac{22\,\zeta_3}{9}-\frac{14\pi^4}{15}   \right)
\\
+&\,C_{S_{ab}}\,n_f\,T_F\left(  -\frac{656}{81}-\frac{50\pi^2}{27} +\frac{8\,\zeta_3}{9} \right) \,. \nonumber
\end{align}
The Casimir operators $C_{S_{ab}}$ for the different channels are 
\begin{align}
C_{S_{q\qb}}&=C_F-\frac{C_A}{2}\,, & C_{S_{qg}}&=\frac{C_A}{2} \,,& C_{S_{gg}}&=\frac{C_A}{2}\,.
\end{align}

\section{Coefficients of the two-loop threshold cross section\label{app:expansion}}

Here we list the expansion coefficients that appear in the two-loop threshold cross section (\ref{sigmaexpanded}). The one-loop coefficients $p_i^{(1)}$ read
\begin{align}
p_0^{(1)}&=-\frac{\pi^2\,\gamma_0^\text{cusp}}{12}\left(C_J+4C_S\right)+c_1^J+c_1^S+2\,\gamma_0^S\,\ln\frac{\mu}{p_T}+2\,\gamma_0^\text{cusp}\,C_S\,\ln^2\frac{\mu}{p_T}\,, \nonumber \\
p_1^{(1)}&=\gamma_0^J+2\gamma_0^S  +4\,\gamma_0^\text{cusp}\,C_S\,\ln\frac{\mu}{p_T}\,,\\
p_2^{(1)}&=\gamma_0^\text{cusp}\left( C_J+4C_S \right)\,,\nonumber
\end{align}
where the lower index indicates the distribution which the coefficients multiply.
The two-loop coefficients $p_i^{(2)}$ are
\begin{align}
p_0^{(2)}&=-\left(C_J+4C_S\right)\left(\frac{\pi^2}{12}\,(\gamma_1^\text{cusp}+\gamma_0^\text{cusp}\,(c_1^J+c_1^S))+\frac{\gamma_0^\text{cusp}\,\zeta_3 }{3}(\beta_0-3(\gamma_0^J+2\gamma_0^S))\right)\nonumber\\
&+\frac{\pi^4\,(\gamma_0^\text{cusp})^2}{480}\left(C_J+4C_S\right)^2+\frac{\pi^2}{12}\left(\gamma_0^J+2\gamma_0^S \right)\left( \beta_0-(\gamma_0^J+2\gamma_0^S) \right)\nonumber\\
&+\frac{\beta_0}{6}\left( \pi^2\,\gamma_0^S-8\gamma_0^\text{cusp}\,C_S\,\zeta_3  \right)+c_1^J\, c_1^S+c_2^J+c_2^S\nonumber\\
&+\ln\frac{\mu}{p_T}\phantom{^1}\bigg\{-\frac{\pi^2\,\gamma_0^\text{cusp}}{6}\left(\left(C_J+4C_S\right)\,\gamma_0^S+4C_S\left(\gamma_0^J+2\gamma_0^S \right)\right) +2\left(\gamma_1^S-\beta_0 \, c_1^S \right)\nonumber\\
&\phantom{+\left(\ln\frac{\mu}{p_T}\right)\phantom{^1}\bigg\{}\;+\frac{2\gamma_0^\text{cusp}\,C_S}{3}\left( \pi^2\,\beta_0+6 \gamma_0^\text{cusp}\,\zeta_3\left(C_J+4C_S\right)\right) +2\gamma_0^S\left(c_1^J+c_1^S\right)\bigg\}\nonumber\\
&+\ln^2\frac{\mu}{p_T}\bigg\{-\frac{\pi^2\,(\gamma_0^\text{cusp})^2\,C_S}{6}\left(C_J+4C_S\right)+2C_S\,\gamma_0^\text{cusp}\left( c_1^J+c_1^S \right)-2\gamma_0^S\left( \beta_0-\gamma_0^S \right)\nonumber\\
&\phantom{+\left(\ln\frac{\mu}{p_T}\right)^2\bigg\{} \; -\frac{1}{3}C_S\left( 4\pi^2\,(\gamma_0^\text{cusp})^2\,C_S-6\gamma_1^\text{cusp} \right)\bigg\}\nonumber\\
&+\ln^3\frac{\mu}{p_T}\bigg\{\frac{4\gamma_0^\text{cusp}\,C_S}{3}\left( 3\gamma_0^S-\beta_0 \right)\bigg\}\nonumber\\
&+\ln^4\frac{\mu}{p_T}\bigg\{2(\gamma_0^\text{cusp})^2\,C_S^2\bigg\}\, ,
\\
\nonumber\\
p_1^{(2)}&=\frac{\pi^2\,\gamma_0^\text{cusp}}{12}\left(C_J+4C_S \right)\left(\beta_0-3(\gamma_0^J+2\gamma_0^S) \right)+(\gamma_0^\text{cusp})^2\,\zeta_3\left( C_J+4C_S \right)^2\nonumber\\
&+\left( \gamma_0^J+2\gamma_0^S \right)\left(c_1^J+c_1^S  \right)-\beta_0\left(c_1^J+2c_1^S  \right)+\gamma_1^J+2\gamma_1^S+\frac{\pi^2\,\beta_0\,\gamma_0^\text{cusp}\,C_S}{3\nonumber}\\
&+\ln\frac{\mu}{p_T}\phantom{^1}\bigg\{ -\pi^2\,(\gamma_0^\text{cusp})^2\,C_S\left( C_J+4C_S\right)+2\gamma_0^S\left(\gamma_0^J+2\gamma_0^S \right)+4\gamma_0^\text{cusp}\,C_S\left( c_1^J+c_1^S\right)\nonumber\\
&\phantom{+\left(\ln\frac{\mu}{p_T}\right)\phantom{^1}\bigg\{}\; +4\left(C_S\,\gamma_1^\text{cusp}-\beta_0\,\gamma_0^S \right)\bigg\}\nonumber\\
&+\ln^2\frac{\mu}{p_T}\bigg\{2\gamma_0^\text{cusp}\,C_S\left( \gamma_0^J+6\gamma_0^S-2\beta_0  \right) \bigg\}\nonumber\\
&+\ln^3\frac{\mu}{p_T}\bigg\{8(\gamma_0^\text{cusp})^2\,C_S^2\bigg\}\, ,
\\
\nonumber\\
p_2^{(2)}&=-\frac{\pi^2\,(\gamma_0^\text{cusp})^2}{4}\left(C_J+4C_S \right)^2+\left(C_J+4C_S \right)\left(\gamma_0^\text{cusp}\,(c_1^J+c_1^S)+\gamma_1^\text{cusp}  \right)+\left( \gamma_0^J+2\gamma_0^S \right)^2\nonumber\\
&-\beta_0\left( \gamma_0^J+4\gamma_0^S\right)\nonumber\\
&+\ln\frac{\mu}{p_T}\phantom{^1}\bigg\{  2\gamma_0^\text{cusp}\,\gamma_0^S  \left(C_J+4C_S \right)+8\gamma_0^\text{cusp}\,C_S\left(\gamma_0^J+2\gamma_0^S \right)-8\beta_0\,\gamma_0^\text{cusp}\,C_S\bigg\} \nonumber \\
&+\ln^2\frac{\mu}{p_T}\bigg\{ 2(\gamma_0^\text{cusp})^2\,C_S\left( C_J+12C_S \right)  \bigg\}\, ,
\\
\nonumber\\
p_3^{(2)}&= \frac{3\gamma_0^\text{cusp}}{2}\left(C_J+4C_S\right)\left(\gamma_0^J+2\gamma_0^S\right)   -\frac{\beta_0\,\gamma_0^\text{cusp}}{2} \left(C_J+8C_S\right) \nonumber \\
&+\ln\frac{\mu}{p_T}\phantom{^1}\bigg\{  6(\gamma_0^\text{cusp})^2\, C_S\left(C_J+4C_S\right)\bigg\} \,,\\
\nonumber\\
p_4^{(2)}&=\frac{(\gamma_0^\text{cusp})^2}{2}\left(C_J+4C_S\right)^2\,.
\end{align}

\end{appendix}


\end{document}